\documentclass[10pt,aps,prd,twocolumn,floats,floatfix,showpacs,superscriptaddress,nofootinbib]{revtex4-2}
\usepackage[utf8]{inputenc} 

\usepackage{graphicx,mathtools,amssymb,amsmath,amsthm,amsfonts,epsfig,epsf,bm,float,subfig,multirow}
\usepackage[outdir=./]{epstopdf}
\usepackage[linktocpage]{hyperref}
\hypersetup{hidelinks}
\hypersetup{pdfstartview=}
\hypersetup{
    colorlinks=true,
    linkcolor=blue,
    filecolor=magenta,      
    urlcolor=BrickRed,
    citecolor=BrickRed,
}
\usepackage[usenames]{color}
\usepackage{tensor}
\usepackage{csquotes}
\usepackage{mathrsfs}
\usepackage{autobreak}

\usepackage[font=small,labelfont=bf, justification=Justified,
   format=plain]{caption}

\usepackage{slashed}

\def\be{\begin{equation}}
\def\ee{\end{equation}}
\def\beq{\begin{eqnarray}}
\def\eeq{\end{eqnarray}}

\usepackage{pifont}

\usepackage[dvipsnames]{xcolor}

\begin{document}

\title{Neutron stars and the cosmological constant problem}

\author{\textbf{Giulia Ventagli}}
\email{ventagli@fzu.cz}
\affiliation{CEICO, Institute of Physics of the Czech Academy of Sciences, Na Slovance 2, 182 21 Praha 8, Czechia}

\author{\textbf{Pedro G. S. Fernandes}}
\email{pgsfernandes@sdu.dk}
\affiliation{CP3-Origins, University of Southern Denmark, Campusvej 55, DK-5230 Odense M, Denmark}
\affiliation{Institut f\"ur Theoretische Physik, Universit\"at Heidelberg, Philosophenweg 16, 69120 Heidelberg, Germany}

\author{\\ \textbf{Andrea Maselli}}
\email{andrea.maselli@gssi.it}
\affiliation{Gran Sasso Science Institute (GSSI), I-67100 L’Aquila, Italy}
\affiliation{INFN, Laboratori Nazionali del Gran Sasso, I-67100 Assergi, Italy}

\author{\textbf{Antonio Padilla}}
\email{Antonio.Padilla@nottingham.ac.uk}
\affiliation{Nottingham Centre of Gravity, University of Nottingham,
University Park, Nottingham NG7 2RD, United Kingdom}
\affiliation{School of Physics and Astronomy, University of Nottingham, University Park, Nottingham NG7 2RD, United Kingdom}

\author{\textbf{Thomas P. Sotiriou}}
\email{Thomas.Sotiriou@nottingham.ac.uk}
\affiliation{Nottingham Centre of Gravity, University of Nottingham,
University Park, Nottingham NG7 2RD, United Kingdom}
\affiliation{School of Mathematical Sciences, University of Nottingham,
University Park, Nottingham NG7 2RD, United Kingdom}
\affiliation{School of Physics and Astronomy, University of Nottingham,
University Park, Nottingham NG7 2RD, United Kingdom}


\begin{abstract}
Phase transitions can play an important role in the cosmological constant problem, allowing the underlying vacuum energy, and therefore the value of the cosmological constant, to change. Deep within the core of neutron stars, the local pressure may be sufficiently high to trigger the QCD phase transition, thus generating a shift in the value of the cosmological constant. The gravitational effects of such a transition should then be imprinted on the properties of the star. Working in the framework of General Relativity, we provide a new model of the stellar interior, allowing for a QCD and a vacuum energy phase transition. We determine the impact of a vacuum energy jump on mass-radius relations,  tidal deformability-radius relations, I-Love-Q relations and on the combined tidal deformability measured in neutron star binaries.  
\end{abstract}

\maketitle

\section{Introduction}\label{sec:Introduction}
Does the vacuum energy gravitate? Within the framework of General Relativity, vacuum energy is expected to gravitate like a cosmological constant, with an equation of state (EOS) $\omega_\Lambda =-1$ giving rise to an accelerated cosmological expansion on large scales \cite{1917KNAB...19.1217D}. Furthermore, observations of distant supernova \cite{SupernovaSearchTeam:1998fmf} and the cosmic microwave background radiation \cite{Planck:2018vyg}, indicate that our universe is undergoing a phase of accelerated expansion, consistent with an underlying cosmological constant of energy density $\rho_\Lambda \sim (\text{meV})^4$. Unfortunately, this observation is wildly out of sync with our theoretical estimates based on naturalness. Indeed, when we apply standard quantum field theory techniques, we find that radiative corrections to the vacuum energy are extremely sensitive to ultra-violet physics, scaling like the fourth power of the field theory cut-off. For a TeV scale cut-off, the expected value of the cosmological constant is at least sixty orders of magnitude greater than the observed value. This is a conundrum known as the cosmological constant problem, widely regarded as one of the most important problems in fundamental physics (for reviews, see \cite{Weinberg:1988cp,Burgess:2013ara,Padilla:2015aaa,Bernardo:2022cck}). 

Theoretical considerations can help guide us in finding the solution to the cosmological constant problem, often in the form of so-called no-go theorems (see, for example, \cite{Weinberg:1988cp,Niedermann:2017cel}). However, there is no proposed solution that enjoys universal support, although a range of interesting ideas span the literature including dynamical neutralisation mechanisms \cite{Brown:1988kg,Liu:2023vqp,Liu:2024blx}, modifications of gravity (both local \cite{Charmousis:2011bf} and  global \cite{Kaloper:2013zca}), and, of course, anthropic considerations \cite{Polchinski:2006gy}, to name just a few. The merits of one model over another are often measured on highly subjective aesthetic grounds, alongside orthogonal phenomenological tests. Unfortunately, direct experimental probes of the solution to the cosmological constant problem have not been extensively explored.

An important aspect of the cosmological constant problem is the role of phase transitions, where the underlying vacuum energy, and therefore the cosmological constant, can change. These could include the QCD and the electroweak phase transitions which are expected to have occurred in the early universe. This makes the fine tuning problem even more severe as we require the final value of the cosmological constant to be tiny compared to the scale of the transitions. 

However, phase transitions may also offer us an opportunity to probe the gravitational effect of vacuum energy directly, offering a tantalising window into the cosmological constant problem. Indeed, deep within the core of neutron stars, the local pressure may be sufficiently high to trigger the QCD phase transition. In the standard scenario, this would be expected to generate a shift in the underlying cosmological constant of order $\lambda_\text{QCD}^4$ where $\lambda_\text{QCD} \sim 200 \text{ MeV}$ is the QCD scale. The gravitational effects of such a transition should be imprinted on the properties of the neutron star, including its mass-radius relations and the tidal deformability \cite{Ozel:2010fw,Steiner:2010fz,Baiotti:2016qnr}. 
For this reason, the possibility of using neutron stars to probe the cosmological constant problem was initiated in \cite{Bellazzini:2015wva,Csaki:2018fls} (see \cite{Kamiab:2011am} for earlier related ideas).

In this paper, we begin a thorough exploration of neutron stars as a novel probe of the cosmological constant problem. Working in the framework of General Relativity, we examine the impact of a vacuum energy phase transition on the properties of the star. We model the interior of the star at low density using a tabulated equation of state (EOS) (specifically SLy~\cite{Douchin:2001sv} and AP4~\cite{Akmal:1998cf}). When the local density exceeds twice the nuclear saturation density, we model the QCD phase using a speed of sound parametrisation advocated in \cite{Tews:2018iwm}. This ensures a continuous subluminal speed of sound and allows us to scan over a large number of semi-realistic profiles in the core of the star, while reducing correlations between effects coming from the QCD and vacuum energy transitions. When the interior pressure exceeds the QCD scale, we include a final transition, allowing the vacuum energy to jump by a constant amount. Although this jump is expected to be positive, going as $\lambda_\text{QCD}^4$, we allow for a range of positive and negative values in order to assess the impact of vacuum energy on the properties of the star. This is not just a mathematical curiosity - one could imagine a hitherto unknown particle physics mechanism than screens some or all of the underlying change in vacuum energy. It is important to ask if neutron stars have the potential to probe whether or not such screening occurs.

Our analysis differs from the original work \cite{Bellazzini:2015wva,Csaki:2018fls} in several important ways. As we have already stated, we use realistic models of dense matter based on nuclear calculations 
instead of piecewise polytropes for the low density EOS, and include an explicit QCD phase prior to the vacuum energy transition modelled using the speed of sound parametrisation. The latter also allows us to fix the EOS of the baseline fluid {\it independently} of the vacuum energy jump. This is in contrast to \cite{Csaki:2018fls} where the energy density of the fluid excitation is forced to increase linearly with the scale of the jump in vacuum energy, tying it directly to the EOS in the innermost layer. Our approach does not have this restriction and we can probe the cosmological constant as an independent parameter. We also extend the analysis to include stellar rotation.  Note that, following a different approach from ours, neutron stars with a dark-energy core were studied in Ref.~\cite{Pretel:2024tjw}, where they considered hybrid stars composed by an ordinary-matter crust and a dark-energy core described by the Chaplygin-type EOS.

It turns out that vacuum energy phase transitions deep inside the core of the star do impact the mass-radius relations: when the vacuum energy contribution is positive, the maximum mass is generically lowered; when the vacuum energy is negative, the maximum mass is generically raised. This differs from \cite{Csaki:2018fls} where the maximum mass is lowered in all cases. This is because we allow the {\it total} energy density to fall inside the core when the vacuum energy changes by a negative amount, in contrast to \cite{Csaki:2018fls}, where the total energy density is always assumed to increase. We have also observed that this behaviour is qualitatively the same for rotating stars.

Interestingly, a large negative value of the vacuum energy jump in the core impacts on the tidal deformability-radius relations, hinting at a new possible family of EOSs, with different properties. Overall, the presence of both the QCD and the vacuum energy phase transitions introduces modifications to the combined tidal deformability-mass relations for neutron stars binary with respect to the vanilla EOSs. Crucially, our results are still in agreement with current constraints from gravitational wave observations~\cite{LIGOScientific:2018hze}.

Finally, we investigate the I-Love-Q universal relations confirming that they still hold for both slowly and fast rotating stars with $\sim 1\%$ accuracy.

The rest of this paper is organised as follows: in section \ref{sec:setup} we describe the standard set-up for both static and rotating neutron  stars, including relevant formulae for the tidal deformability,  moment of inertia and other important quantities. In section \ref{sec:EOS}, we describe our EOS modelling in some detail, including the transitions to a QCD phase as well as the vacuum energy transition. We explain the differences and similarities with other approaches, in particular \cite{Csaki:2018fls}. In section \ref{sec:Impl}, we run through our numerical implementation and present our results in detail in section \ref{sec:results}. Finally, in section \ref{sec:conc}, we conclude.

\section{Setup} \label{sec:setup}
%
\subsection{Static Neutron Stars}\label{Sec:static}
%
We first discuss the case of a static, spherically symmetric star, described by the metric
\be\label{eq:eqconfig}
    ds^2 = g^{(0)}_{\mu\nu} dx^\mu dx^\nu = -e^{\nu(r)} dt^2 + e^{\mu(r)} dr^2 + r^2 d\Omega^2\ .
\ee
We assume matter to be described as a perfect fluid with a stress-energy tensor given by
\be
    T_{\mu\nu}^{(0)} = (\epsilon + p) u_\mu u_\nu + p\,g^{(0)}_{\mu\nu},
    \label{eq:stress-energy}
\ee
where $u_\mu=(-e^{\nu/2},0,0,0)$ is the fluid's four velocity, $p$ is the pressure and $\epsilon$ is the energy density. The latter are related to each other by a barotropic EOS, $\epsilon = \epsilon(p)$. 
Details on the specific form of the EOS we use to model the neutron star structure are provided in Section~\ref{sec:EOS}. The metric functions $\nu$ and $\mu$, and the pressure $p$ are determined as solutions of the Tolman-Oppenheimer-Volkoff (TOV) system of equations~\cite{Tolman:1939jz,Oppenheimer:1939ne}
\begin{align}
    \label{eq:tov1}
    m'(r) &= 4 \pi r^2 \epsilon(r)\ ,\\
    \label{eq:tov2}
    p'(r) &= -\frac{p(r)+\epsilon(r)}{r[r-2 m(r)]} G [m(r)+4\pi r^3 p(r)]
     ,\\
    \label{eq:tov3}
    \nu'(r) &= -\frac{2p'(r)}{p(r)+\epsilon(r)}\ ,
\end{align}
where primes denote differentiation with respect to the radial coordinate $r$. In the rest of this paper, we will consider vacuum configurations outside the star, where $T_{\mu\nu}=0$ and the spacetime is asymptotically flat. Equations~\eqref{eq:tov1}-\eqref{eq:tov3} can be solved by imposing suitable boundary conditions at the center of the star $(r=r_0)$ and integrating forward to the surface $(r=R)$ where the pressure vanishes and $m(R)=M$, with $M$ and $R$ being the neutron star's gravitational mass and radius, respectively. We work in units where $c=1$, $G=1$ and $M_\odot=1$, unless stated otherwise.

%
\subsubsection{Tidal deformations}\label{Sec:tidal}
%
In this section we summarize the basic ingredients required to compute relativistic tidal deformations of compact objects, focusing on the dominant, $l=2$, quadrupolar contribution (see~\cite{Hinderer:2007mb,Baiotti:2019sew,Chatziioannou:2020pqz} 
for detailed reviews).

We consider a neutron star immersed in an external static and quadrupolar tidal field $\mathcal{E}_{ij}$. At the leading order, the quadrupole moment, $Q_{ij}$, induced on the star is linearly proportional to $\mathcal{E}_{ij}$. 
The coupling between such quantities is given by the dimensionful tidal deformability $\lambda$
\be
    Q_{ij} = -\lambda \mathcal{E}_{ij}=-\frac{2}{3}k_2R^5\mathcal{E}_{ij}\ ,\label{eq:adiabaticapprox}
\ee
where $k_2$ is the dimensionless apsidal Love number \cite{Thorne:1997kt,Hinderer:2007mb,Hinderer:2009ca,Binnington:2009bb,Damour:2009vw}

To compute $k_2$ (or equivalently $\lambda$) we consider linear perturbations of the background solution~\eqref{eq:eqconfig}
\be
    g_{\mu\nu} = g^{(0)}_{\mu\nu} + h_{\mu\nu}\ ,
\ee
where $h_{\mu\nu}$ depends on $(r,\theta,\phi)$. The symmetry of $g_{\mu\nu}^{(0)}$ allows to expand the perturbations in spherical harmonics $Y_{lm}(\theta,\phi)$, decoupling them into axial and polar modes, according to their transformation rules under parity.\footnote{For spherically symmetric backgrounds, axial and polar sectors disentangle and can be treated separately.} The quadrupolar contribution we consider requires calculations of the polar sector only. Without loss of generality, we fix $m=0$, due to the axial symmetry of the tidal field, such that the ($l=2$) metric perturbations read
\begin{align}
 h_{\mu\nu} = & e^{\nu(r)} H(r) Y_{20}(\theta,\phi) dt^2 + e^{\mu(r)} H(r) Y_{20}(\theta,\phi) dr^2 \nonumber\\
 &+ r^2 K(r) Y_{20}(\theta,\phi) d\Omega^2\ .\label{eq:metrich}
\end{align}
In the same spirit we introduce linear perturbations of the stress-energy tensor
\begin{equation}
T_{\mu\nu}=T_{\mu\nu}^{(0)}+\delta T_{\mu\nu}\ .
\end{equation}
Expanding $\epsilon$, $p$ and the fluid velocity around their background values, we get
\begin{equation}
\delta T^0{_0}=-\delta \epsilon Y_{20}(\theta,\phi)\quad\ ,\quad 
\delta T^i{_i}=\delta p Y_{20}(\theta,\phi)\ .
\end{equation}
The Einstein field equations allow us to express $K(r)$ as a function of $H(r)$, $K'(r) = H' + H(r) \nu'(r)$,  casting the perturbations in terms of a single second order differential equation
\begin{multline}\label{eq:diffH}
    H''(r) -\frac{2}{r} e^\mu \big[ 2\pi G r^2 (\epsilon-p)-1\big]H'(r) \\
    -2 e^\mu \bigg\{ \frac{3}{r^2} -2\pi G [5\epsilon +9 p + (\epsilon+p) f ] \\
    +2 G^2 e^\mu \left( \frac{m(r)}{r^2}+4\pi r p \right)^2 \bigg\} H(r) = 0\ ,
\end{multline}
where $f=d\epsilon/dp$. Note that we have restored $G$ in the equations above.

Equation~\eqref{eq:diffH} can be further reduced to a coupled first order system of ODEs, by introducing $\beta(r) = dH/dr$. We integrate this system from the stellar center, with boundary conditions
\begin{equation}
 H(r_0) \sim a_0 r_0^2 \quad\ ,\quad
\beta(r_0) = 2 a_0 r_0\ ,
\end{equation}
where $a_0$ is a numerical constant\footnote{The value of this constant is irrelevant as it cancels out in the final calculation of $k_2$.}, up to the surface where $H(r)$ is matched to the vacuum solution, and $T_{\mu\nu}=0$. Outside the star $H(r)$ can be expressed analytically in terms of the associated Legendre functions ${\cal Q}_{22}$ and ${\cal P}_{22}$~\cite{Thorne:1967,Hinderer:2007mb,Olver:2010}
\begin{equation}
H(r\ge R)=c_{1} {\cal P}_{22}\left(r/M-1\right)+
c_{2} {\cal Q}_{22}\left(r/M-1\right)\ ,\label{eq:HrR}
\end{equation}
where $c_{1,2}$ are two constants of integration found through the matching. Finally, we plug Eq.~\eqref{eq:HrR} into the metric, and expand the $g_{tt}$ component at spatial infinity, which can be written as the sum of a decaying ($Q^2_2(r/M-1)\sim r^{-3}$ ) and a growing ($P^2_2(r/M-1)\sim r^2$) mode. To extract the star multipole moments we match this expression with the asymptotic form of $g_{tt}$\footnote{We write the metric, and hence $g_{tt}$, in the so-called ACMC coordinates~\cite{Thorne:1980ru}.}
\begin{multline}\label{eq:metrExp}
    -\frac{1+g_{tt}}{2} = -\frac{M}{r}-\frac{3}{2}\frac{Q_{ij}}{r^3}n^in^j+...+\frac{\mathcal{E}_{ij}}{2}r^2 n^i n^j+...\ ,
\end{multline}
where $n^i=x^i/r$ \cite{Thorne:1980ru}. This procedure allows us to identify the growing (tidal field) and decaying (quadrupole deformation) solutions which can be related to the two integration constants (i.e. with the value of $H(R))$. We can finally express the Love number as a function of $H(r)$ and its first derivative at the surface
\begin{multline}
    k_2 = \frac{8 C^5}{5}(1 - 2C)^2(2 + 2C(y-1)-y)\\
    \times\bigg\{ 2 C \big[6 - 3y + 3C(5y - 8)\big] \\
    + 4C^3 \big[13 - 11 y + C(3y - 2) + 2C^2(1 + y)\big] \\
    + 3 (1 - 2C)^2 \big[ 2 - y + 2C(y - 1) \big] \text{ln}(1 - 2C) \bigg\}^{-1},
\end{multline}
where $C=M/R$ is the stellar compactness and $y=R H'(R)/H(R)$. 

In Section~\ref{sec:results}, when discussing the neutron star properties, we will often show results in terms of the dimensionless tidal deformability
\be\label{eq:dimlesstidal}
\bar{\lambda} = \frac{2}{3} \frac{k_2}{C^5} = \frac{\lambda}{M^5}\ ,
\ee
as is commonly done in the literature.

%
\subsection{Slowly Rotating Neutron Stars}\label{sec:statstars}
%
We construct slowly rotating neutron stars in the Hartle-Thorne approach~\cite{Hartle:1967he,Hartle:1968si}. 
We consider a stationary and axisymmetric metric that include up to quadratic corrections in the star rotation
\begin{equation}
    \begin{aligned}
            ds^2 =& - e^{\bar{\nu}(\bar{r})} \left[ 1 + 2\bar{h}_2(\bar{r})P_{2}(\cos\theta) \right] dt^2 \\
            &+ e^{\bar{\lambda}(\bar{r})} \left[1+\frac{2 \bar{m}_2(\bar{r})P_{2}(\cos\theta)}{r-2\bar{m}(\bar{r})} \right] d\bar{r}^2\\
            & +\bar{r}^2 \left[ 1+2 \bar{K}_2(\bar{r})P_{2}(\cos\theta) \right]\\
            &\times \left\{ d\theta^2 +\text{sin}^2\theta \left\{ d\phi -\left[ \Omega_\ast -\bar{\omega}_1(\bar{r}) P'_1(\text{cos}\theta)  \right]dt \right\}^2 \right\}\ ,
    \end{aligned}
\end{equation}
where $\bar{m}(\bar{r})=(1-e^{-\bar{\lambda}(\bar{r})})\bar{r}/2$, $\Omega_\ast$ is the constant angular velocity of the star, $P_\textit{l}(\text{cos}\theta)$ are the Legendre polynomials of \textit{l}-th order and $P'_1=dP_1/d(\text{cos}\theta)$. Linear contributions in the spin arise from $\bar{\omega}_1$, while second order terms are determined by $(\bar{m}_2,\bar{K}_2,\bar{h}_2)$. 

For the perturbative scheme to hold, we transform the radial coordinate by~\cite{Hartle:1967he,Yagi:2013awa,Yagi:2013mbt}
\begin{equation}
    \bar{r}(r,\theta)=r+\eta^2 \xi_2(r) P_{2}(\cos\theta) \ ,
\end{equation}
where $\xi_2(r)$ is chosen such that the energy density in the new coordinates is identical to the unperturbed one, that is
\begin{equation}
    \epsilon[\bar{r}(r,\theta,\phi)]=\epsilon(r)=\epsilon^{(0)}(r).
\end{equation}
All functions in the new coordinate system will be identified by non-barred quantities, e.g. $\bar{\nu}(\bar{r})\to\nu(r)$.

The four-velocity of the fluid, assuming the stress-energy tensor~\eqref{eq:stress-energy}, is now given by
\begin{equation}
    u^\mu =(u^0,0,0,\eta\, \Omega_\ast u^0)\ .
\end{equation}
At linear order in the rotation, the function $\omega_1$ is determined by the $(t,\phi)$ component of the Einstein equations, which yields
\begin{equation}\label{eq:slowly1}
    \frac{d^2\omega_1}{dr^2}+4\frac{1-\pi r^2(\epsilon+p)e^\lambda}{r}\frac{d\omega_1}{dr}-16\pi(\epsilon+p)e^\lambda\omega_1=0.
\end{equation}
The interior solution of Eq.~\eqref{eq:slowly1} can be solved by imposing suitable boundary conditions on $\omega_1$ at the centre of the star and integrating up to the surface. One can then retrieve the exterior solution by integrating Eq.~\eqref{eq:slowly1} in the absence of matter (i.e. $\epsilon=0$ and $p=0$), modulo an integration constant. The latter can be fixed by matching the exterior and interior solution at the $r=R$. Outside the star $\omega_1$ is given by
\begin{equation}
    \omega^{\text{ext}}_1=\Omega_\ast-\frac{2S}{r^3}.
\end{equation}
where $S$ can be identified with the total angular momentum of the star. The moment of inertia $I$ can then be computed as
\begin{equation}
    I\equiv\frac{S}{\Omega_\ast}.
\end{equation}
At the second order in the rotation, the field's equations for the metric functions $(m_2,K_2,h_2)$ are given by
\begin{widetext}
\begin{align}
     m_2 =& -r e^{-\lambda}h_2 + \frac{1}{6} r^4 e^{-(\nu +\lambda)} \left\{ r e^{-\lambda} \left(\frac{d\omega_1}{dr}\right)^2 + 16 \pi r \omega_1^2(\epsilon+p)\right\}\ ,\label{eq:slowly2.1} \\ 
     \frac{dK_2}{dr} =& -\frac{dh_2}{dr} + \frac{r-3m-4\pi p r^3}{r^2}e^\lambda h_2 + \frac{r-m+4 \pi p r^3}{r^3}e^{2\lambda}m_2\ , \label{eq:slowly2.2} \\ 
     \frac{dh_2}{dr} =& -\frac{r-m+4\pi p r^3}{r}e^\lambda \frac{dK_2}{dr} + \frac{3-4\pi(\epsilon+p)r^2}{r}e^\lambda h_2+ \frac{2}{r}e^\lambda K_2 +\frac{1+8\pi p r^2}{r^2}e^{2\lambda}m_2 +\frac{r^3}{12}e^{-\nu}\left(\frac{d\omega_1}{dr}\right)^2\nonumber \\
     &-\frac{4\pi(\epsilon+p)r^4\omega_1^2}{3r}e^{-\nu+\lambda}\ , \label{eq:slowly2.3}
\end{align}
\end{widetext}
while the $\theta$-component of $\nabla^\mu T_{\mu\nu}=0$ yields
\begin{equation}\label{eq:slowlydT}
    \xi_2 = -\frac{r^2 e^{-\lambda}(3 h_2 + e^{-\nu}r^2\omega_1^2)}{3(m+4\pi p r^3)}\ .
\end{equation}
Once again, to solve Eqs.~\eqref{eq:slowly2.1}-\eqref{eq:slowlydT} we need to study the interior and the exterior problem individually, and match the two solutions at the stellar surface~\cite{Yagi:2013awa} (the values of the metric functions in vacuum can be obtained analytically, 
requiring asymptotic flatness at spatial infinity, 
and are shown in Appendix~\ref{App:Slowly}). Finally, the spin-induced quadrupole moment $Q$ can be determined from the coefficient proportional to $P_2(\text{cos}\theta)/r^3$ in the exterior solution for the $g_{tt}$ metric component, and is given by
\begin{equation}
    Q = -\frac{S^2}{M}-\frac{8}{5}A M^3\ ,
\end{equation}
where $A$ is an integration constant obtained through the matching procedure.

%
\subsection{Fast Rotating Neutron Stars}\label{sec:rotstars}
%

To construct fast rotating neutron stars solutions we follow closely the method developed in Ref.~\cite{Komatsu:1989zz}, assuming uniform rotation. We employ a stationary and axisymmetric metric ansatz in quasi-isotropic coordinates
\begin{equation}
    \begin{aligned}
            ds^2 =& - e^{\gamma + \rho} dt^2 + e^{\gamma-\rho} r^2 \sin^2\theta \left(d\phi - \omega dt\right)^2 \\&+ e^{2\alpha} \left(dr^2 + r^2 d\theta^2\right)\ ,
    \end{aligned}
\end{equation}
where the potentials $(\gamma,\rho,\omega,\alpha)$ depend only on the coordinates $r$ and $\theta$. We assume once again that matter is described by a perfect fluid with the same stress-energy tensor given in Eq.~\eqref{eq:stress-energy}. The four-velocity is now expressed as
\begin{equation}
    u^\mu = \frac{e^{-(\gamma+\rho)/2}}{\sqrt{1-v^2}} \left( 1,0,0,\Omega \right)\ ,
\end{equation}
where $\Omega$ is the angular velocity of matter measured from infinity and 
$v = \left(\Omega-\omega\right)r\sin \theta e^{-\rho}$ is the proper velocity with respect to the zero angular momentum observer. In this setup the Einstein's field equations take the form~\cite{Komatsu:1989zz, Friedman:2013xza}
\begin{equation}
    \nabla^2 \left[ \rho e^{\gamma/2} \right] = S_\rho (r, \theta)\ ,
    \label{Eq:FEq1}
\end{equation}
\begin{equation}
    \nabla^2 \left[ r \sin \theta \cos \phi \omega e^{(\gamma-2\rho)/2} \right] = r \sin \theta \cos \phi S_\omega (r, \theta)\ ,
    \label{Eq:FEq2}
\end{equation}
\begin{equation}
    \left(\partial_{\varpi}^2 + \partial_z^2\right) \left[ \varpi \gamma e^{\gamma/2} \right] = \varpi S_\gamma (r, \theta)\ ,
    \label{Eq:FEq3}
\end{equation}
\begin{equation}
    \partial_\theta \alpha = S_\alpha\left( r,\theta \right)\ ,
    \label{Eq:FEq4}
\end{equation}
where $\nabla^2$ is the Laplacian operator in 3D spherical coordinates, $\varpi = r \sin \theta$, $z = r \cos \theta$ and the source terms $S_\rho$, $S_\omega$, $S_\gamma$ and $S_\alpha$ are shown in Appendix~\ref{Appendix:Rotating}. Note that the differential operator appearing in Eq.~\eqref{Eq:FEq3} is corresponds to the 2D Laplacian in Cartesian coordinates $(\varpi,z)$.

The elliptic character of the field equations allows us to recast Eqs.~\eqref{Eq:FEq1}-\eqref{Eq:FEq2} in integral form
\begin{equation}
    \rho = -\frac{e^{\gamma/2}}{4\pi} \int_0^\infty dr' \int_0^\pi d\theta' \sin \theta' \int_0^{2\pi} d\phi' r'^2 \frac{S_\rho\left(r', \theta'\right)}{|\mathbf{r}-\mathbf{r'}|},
    \label{Eq:FEq1Int}
\end{equation}
\begin{equation}
\begin{aligned}
        \omega =& -\frac{e^{(2\rho-\gamma)/2}}{4\pi r \sin\theta \cos \phi} \int_0^\infty dr' \int_0^\pi d\theta' \sin \theta' \\& \int_0^{2\pi} d\phi' r'^3 \cos \phi' \frac{S_\omega\left(r', \theta'\right)}{|\mathbf{r}-\mathbf{r'}|},
\end{aligned}
\label{Eq:FEq2Int}
\end{equation}
\begin{equation}
    \gamma = \frac{e^{-\gamma/2}}{2\pi r \sin \theta} \int_0^\infty dr' \int_0^{2\pi} d\theta' r'^2 \sin \theta' S_\gamma\left(r', \theta'\right) \log |\mathbf{r}-\mathbf{r'}|.
    \label{Eq:FEq3Int}
\end{equation}

The integrals above can be efficiently evaluated using the standard expansions of the Green's functions in angular harmonics. These are given in Appendix~\ref{Appendix:Rotating}. From the metric functions we compute the moment of inertia $I$ and the spin induced quadrupole moment $Q$. The latter is extracted from the asymptotic expansion of the $g_{tt}$ component, following the method described in~\cite{Laarakkers:1997hb}. The moment of inertia is determined as $I=J/\Omega$, where $J$ is the neutron star's angular momentum~\cite{Paschalidis:2016vmz}.

\section{Equation of state models}\label{sec:EOS}

Given the large uncertainties in modelling the behaviour of nuclear matter above the saturation density $\rho_0\simeq2.7 \times 10^{14}\, \text{g}/\text{cm}^{3}$, we assume that the neutron star structure is described by an EOS composed of three regions, described by different approaches.

In the low-density part, up to a threshold density $\rho_\text{tr}=2\rho_0$, we adopt a nucleonic tabulated EOS (region I). For $\rho>\rho_\text{tr}$, within the high-density regime, the nuclear matter undergoes a QCD phase transition, and we employ a phenomenological EOS (region II). Finally, in the inner stellar core, for values of the pressure larger than $(200\, \text{MeV})^4$, we assume that a vacuum energy phase transition occurs and we use, again, a phenomenological EOS (region III).

%
\subsection{Low-density equation of state}\label{sec:lowDens}
%

For the low density part ($\rho\le \rho_\text{tr})$ we choose two models commonly used in literature, namely the SLy~\cite{Douchin:2001sv} and the AP4~\cite{Akmal:1998cf} EOS. Such models are obtained through different methodologies and calculation schemes. Both of them provide soft nuclear matter, and are consistent with current astrophysical observations, either in the electromagnetic or the gravitational wave band~\cite{Miller:2021qha,Miller:2019cac,Riley:2019yda,Riley:2021pdl,LIGOScientific:2020aai,LIGOScientific:2018hze}.

The SLy EOS is based on a mean field theory approach, in which nucleons interact with an effective Skyrme potential~\cite{Chabanat:1997qh,Chabanat:1997un}, while the AP4 EOS is constructed in the context of many body theory, and includes the two- and three-body nucleon interactions based on the Argonne $v_{18}$ and the Urbana IX potentials~\cite{Akmal:1998cf}. Both EOSs predict nuclear matter which include neutrons, protons, electrons and muons in beta equilibrium.

Note that our treatment differs from that used in~\cite{Csaki:2018fls}, where the low-density regime has been modelled through phenomenological piecewise polytropic EOSs. We do this in order to reduce correlations between effects coming from the QCD and vacuum energy phase transitions, and those arising from the wide parameter space allowed by piecewise EOSs.

%
\subsection{Modeling the QCD transition}
%

For $\rho > \rho_\text{tr}$, we extend the low-density EOS using an agnostic speed-of-sound model~\cite{Tews:2018kmu,Tews:2018iwm} (see also Ref.~\cite{Annala:2019puf} for an alternative approach). As shown in~\cite{Tews:2018iwm}, this approach extends the commonly used piecewise polytropic 
parametrisations~\cite{Read:2008iy,Hebeler:2010jx,Hebeler:2013nza,Raithel:2016bux,Annala:2017llu}, with the advantage of keeping $c_s$ continuous across the star. In this approach we construct a linear approximation of the speed of sound $c_s(\rho)$,
\be
    c_s^2 = \frac{\partial p(\epsilon)}{\partial \epsilon}\ ,
\ee
randomly sampling six reference points $(\rho,c_s)$ within $\rho\in(\rho_\text{tr}, 12\,\rho_0)$ and $c_s\in(0, 1)$, and connecting them by linear segments. We assume nuclear matter to respect causality, i.e. we discard models with $c_s > 1$, avoiding unphysical EOSs. Note that by randomly selecting $(\rho,c_s)$ points, the QCD phase could develop phase transitions.

We determine the EOS with the following procedure. We start from the values of the energy, pressure and $c_s$ at $\rho_\text{tr}$ obtained by continuity from the low-density regime. Then, we construct iteratively the energy-pressure relation by choosing a mass-density step $\Delta \rho$, and computing 
\begin{align}
    \rho_{i+1} =& \rho_{i} + \Delta\rho\ , \\
    \epsilon_{i+1} =& \epsilon_{i} + \Delta\epsilon = \epsilon_{i} + \Delta\rho \left( \frac{\epsilon_i + p_i}{\rho_i} \right)\ ,\\
    p_{i+1} =& p_{i} + c_s^2 \Delta\epsilon\ ,
\end{align}
where $i=0$ identifies quantities at the transition density, and to derive the 
second equation we have exploited the thermodynamic relation $p = \rho\,\partial\epsilon/\partial\rho - \epsilon$, valid at zero temperature.

This piece of EOS modelling a QCD phase in the core of the star is absent in Ref.~\cite{Csaki:2018fls}, where the excitation of the baseline fluid energy density is tied directly to the vacuum energy jump. Our choice allows us to introduce a new matter phase in the innermost layer of the star, while also fixing the EOS of the baseline fluid independently of the vacuum energy shift.

\subsection{Vacuum energy phase transition in the core}\label{sec:CC}

As a last step, we implement the vacuum energy shift in the core of the star. We assume that this phase transition is triggered whenever a certain threshold in pressure $p_c$ is crossed, which we choose to be proportional to the QCD scale, that is $p_c = (200\,\text{MeV})^4$. As a consequence, only neutron stars that are sufficiently dense will develop a jump in vacuum energy.

As we have mentioned in Section~\ref{sec:Introduction}, this shift in energy can be interpreted as a new effective cosmological constant term $\Lambda$ for the new exotic phase. In this region, the total pressure, energy density and mass density of the star will then be a sum of the fluid and the cosmological constant contributions. 
Additionally, the Israel junction conditions~\cite{Israel:1966rt} requires that the (total) pressure is continuous, leading to a jump in (total) energy density and (total) mass density. We then have
\beq
    \label{eq:pTotal}
    p_\text{t} =  
    \begin{cases}
    p_\text{fl} &  p_\text{t}< p_c\\
    p_\text{fl} - \Lambda & p_\text{t}\geq p_c
    \end{cases}\\
    \label{eq:epsTotal}
    \epsilon_\text{t} =  
    \begin{cases}
    \epsilon_\text{fl} &  p_\text{t}< p_c\\
    \epsilon_\text{fl} + \Lambda & p_\text{t}\geq p_c
    \end{cases}\\
    \label{eq:rhoTotal}
    \rho_\text{t} =  
    \begin{cases}
    \rho_\text{fl} &  p_\text{t}< p_c\\
    \rho_\text{in} \, \text{exp}\left( \int_{\epsilon\ast}^{\epsilon_\text{t}} \frac{d\tilde{\epsilon}_\text{t}}{\tilde{\epsilon}_\text{t} + p_\text{t}}\right) & p_\text{t}\geq p_c
    \end{cases}
\eeq
where a subscript `t' indicates total quantities, a subscript `fl' indicates quantities related to the baseline fluid. Here we have derived the expression for the total mass density after the transition by employing the thermodynamic law $\text{d}(\epsilon/\rho)=-p\,\text{d}(1/\rho)$. Finally, $\rho_\text{in}$ is given by
\be\label{eq:rhoIn}
   \rho_\text{in} = \rho_\text{out} \left(\frac{\epsilon_\text{in} + p_c}{\epsilon_\text{out} + p_c}\right)\ ,
\ee
where the label `in' (`out') indicates total quantities at the junction $p_c$ before (after) the $\Lambda$ phase transition. We stress that, for the total pressure, it must hold that $p_\text{in}=p_\text{out}=p_c$. In order to achieve this, the fluid pressure will experience a discontinuity jump proportional to $\Lambda$. We then retrieve the following relations for the total energy density
\begin{align}
    &\epsilon_\text{out} = \epsilon_\text{fl}(p_c)\ ,\\
    &\epsilon_\text{in} = \epsilon_\text{fl}(p_c + \Lambda) + \Lambda\ ,
\end{align}
where $\epsilon_\text{fl} = \epsilon_\text{fl}(p_\text{fl})$ is the baseline EOS we derived in the previous sections. Using Eqs.~\eqref{eq:pTotal}-\eqref{eq:rhoTotal}, we can construct our final tabulated EOSs.

Prior to the vacuum energy transition, the gluon condensate is thought to contribute negatively to the overall vacuum energy by an amount $-0.0034~(\text{GeV})^4$ \cite{Donoghue:2017vvl}. This is the standard value usually quoted in the literature, despite some mild controversy \cite{Holdom:2007gg,Holdom:2009ma}. The controversy stems from the fact that the corresponding operator is divergent in perturbation theory and so this is really a subtracted result. There is a smaller contribution from light quark masses that is rigorously calculable, going as $-F_\pi^2 m_\pi^2$, where $F_\pi=92$ MeV is the pion decay constant and $m_\pi=135$ MeV is the pion mass \cite{Donoghue:2016tjk}. Regardless of the precise details, it is reasonable to expect the vacuum energy to {\it increase} by $(\mathcal{O}(100) \text{MeV})^4$ as we transition towards the inner core of the neutron star.\footnote{We thank John Donoghue for educating us on these points.}

In this work, we choose $\Lambda$ to be in the range $$\big(-(194\, \text{MeV})^4, (194\, \text{MeV})^4\big),$$ including $\Lambda = 0$ (that is, if no phase transition develops in the core). Of course, the discussion of the previous paragraph suggests that $\Lambda$ is expected to be positive and $(\mathcal{O}(100) \text{MeV})^4$ in a standard scenario. Here we allow for more possibilities, in order to thoroughly explore the space of $\Lambda$ and its impact on the properties of the star. We are also open to the possibility that there is some unknown physics at work, potentially screening some or all of the corresponding cosmological constant. Note that the range of values that we can pick for $\Lambda$ is bounded from below, since at the junction we have $p_\text{fl} = p_c + \Lambda$, with $p_c = (200 \,\text{MeV})^4$. For $\Lambda \leq - p_c$, the fluid pressure will be zero or negative, which is clearly not a physical scenario.

As discussed in Section~\ref{sec:Introduction}, our model describing the effect of the vacuum energy is inherently different from the approach used in Ref.~\cite{Csaki:2018fls}. In this previous work, the fluid EOS was modeled differently, being described by piecewise polytropes, and it did not include an intermediate QCD state of matter. Crucially, it was also assumed that the mass density is not modified by a $\Lambda$ contribution, forcing the jump in total energy density to be positive and proportional to $\Lambda$, after consideration on the convexity of free energy of the fluid $\left( \frac{\partial^2F_\text{fl}}{\partial V^2} \right)_\text{T, N} > 0$.

With our approach, we fix the EOS using the speed of sound parametrisation, then allow for a variety of different cosmological constants, both positive and negative. In order for the relevant continuity equations to hold at the transition, we see that the mass density must be modified by the cosmological constant contribution. Depending on whether the mass density increases or decreases, there can be a jump in energy density of either sign, set by the continuity relations. Indeed, we can see this directly from Eq.~\eqref{eq:rhoIn}, which gives
\be
    \frac{\Delta\epsilon}{\rho_\text{in}/\rho_\text{out}-1} = (\epsilon_\text{out} + p_c),
\ee
where $\Delta\epsilon = \epsilon_\text{in} - \epsilon_\text{out}$. The quantity on the right-hand side is positive, assuming the fluid outside the junction satisfies the null energy condition. Hence we find that the sign of $\Delta\epsilon$ depends on $\Delta\rho = \rho_\text{in} - \rho_\text{out}$, that is 
\be\label{eq:condEn}
    \Delta\epsilon > 0 \, \leftrightarrow \, \Delta\rho > 0.
\ee
Furthermore, we show the conditions that arise from the the convexity of the free energy in Appendix~\ref{App:convex}.

\section{Numerical implementation}\label{sec:Impl}

The calculations required to build our EOS following the procedure described in section~\ref{sec:EOS} have been performed using a \texttt{Julia}~\cite{Julia-2017} notebook, which we make publicly available at Ref.~\cite{Ventagli2023}. Note that the EOSs we construct following Eqs.~\eqref{eq:pTotal}-\eqref{eq:rhoTotal} are limited by the maximum and minimum pressure and energy density provided in the table which represent our baseline nuclear model.

When modelling the QCD part of the EOS, we create two different datasets composed of $10^2$ and $10^3$ EOSs, each one given by a 6 random $c_s$ segments. To distinguish between each specific QCD modification, we introduce a bookkeeping parameter `z', that identifies a specific choice of 6 random segments in region II. Each of these new EOSs is matched in the lower density part with either SLy or AP4, leading hence to $2 \times 10^2$ and $2 \times 10^3$ baseline EOSs. Finally, the high density region of the EOSs is connected to a vacuum energy transition for 10 different values of $\Lambda$ in the range $\big(-(194\, \text{MeV})^4, (194\, \text{MeV})^4\big)$. With this procedure we obtain two ensembles with $2 \times 10^3$ and $2 \times 10^4$ EOSs.

For static stars, we first integrate the TOV equations~\eqref{eq:tov1}-\eqref{eq:tov3}, together with the tidal deformability equation~\eqref{eq:diffH} using a $4^\text{th}$ order Runge-Kutta method. The \texttt{Julia} notebook we developed for this task is publicly available at Ref.~\cite{Ventagli2023}. We start our integration from $r_0 \sim 10^{-5}\,\text{km}$ up to distances $r_\text{max}\sim 300\,\text{km}$. For every EOS we integrate the system of equations from a central pressure $p_0$ up to $10^{-12}p_0$, which we consider as \textit{vacuum} condition to identify the stellar radius, where $p(R)=0$. The value of the central pressure $p_0$ varies from $1.38 \times 10^{34}\,\text{g}\,\text{cm}^{-1}\text{s}^{-2},$ up to the highest available pressure in the table, typically around $10^{36}\, \text{g}\,\text{cm}^{-1}\text{s}^{-2}$. Varying $p_0$ we can then construct different stellar configurations corresponding to a given EOS, i.e. to build its mass-radius curve. We require the curves obtained with this procedure to be in agreement with the heaviest neutron star observed so far, PSR J0952-0607~\cite{Romani:2022jhd}, which has a mass $M_\text{max}=(2.35\pm 0.17)M_\odot$. Therefore, we discard EOSs which predict a maximum mass smaller than $M_\text{max}$.

For fast rotating stars, we use the publicly available \texttt{RNS} code~\cite{RNS_Code}, which allows to construct rapidly rotating neutron star 
solutions in General Relativity for a given a tabulated EOS. The code uses the formalism described in Sec.~\ref{sec:rotstars} and the method developed in ~\cite{Komatsu:1989zz} to solve iteratively the integral equations~\eqref{Eq:FEq1Int}-\eqref{Eq:FEq3Int}, together Eq.~\eqref{Eq:FEq4}. We refer the reader to \cite{Friedman:2013xza} for a detailed description 
of such iterative scheme. For slowly rotating stars, 
we have written a \texttt{FORTRAN} code, available 
at~\cite{SGREP_REPO}.

\section{Results} \label{sec:results}

\subsection{Analysis of dataset}\label{sec:dataset}

\begin{figure*}
\begin{center}
\includegraphics[width=0.5\linewidth]{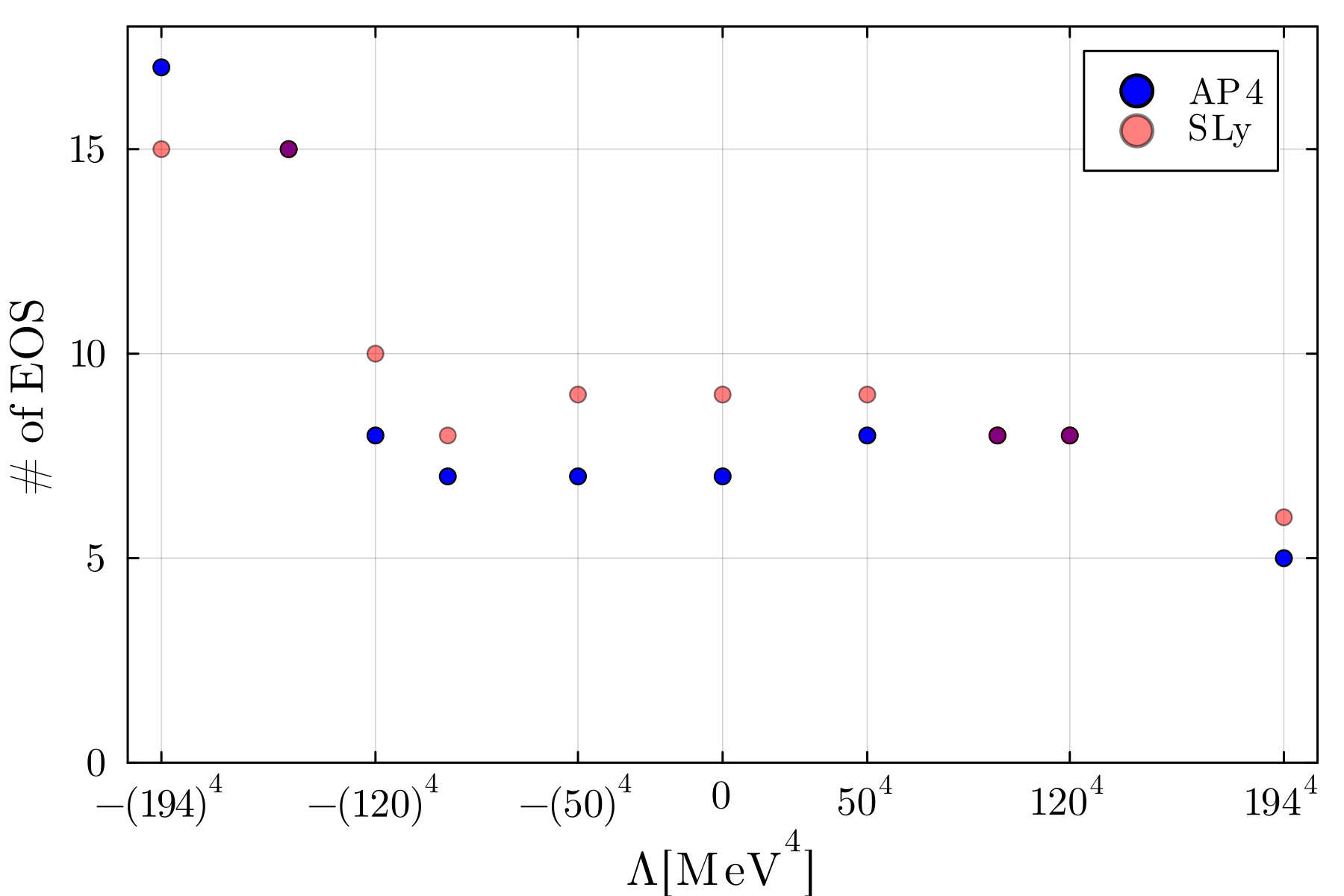}\hfill
\includegraphics[width=0.5\linewidth]{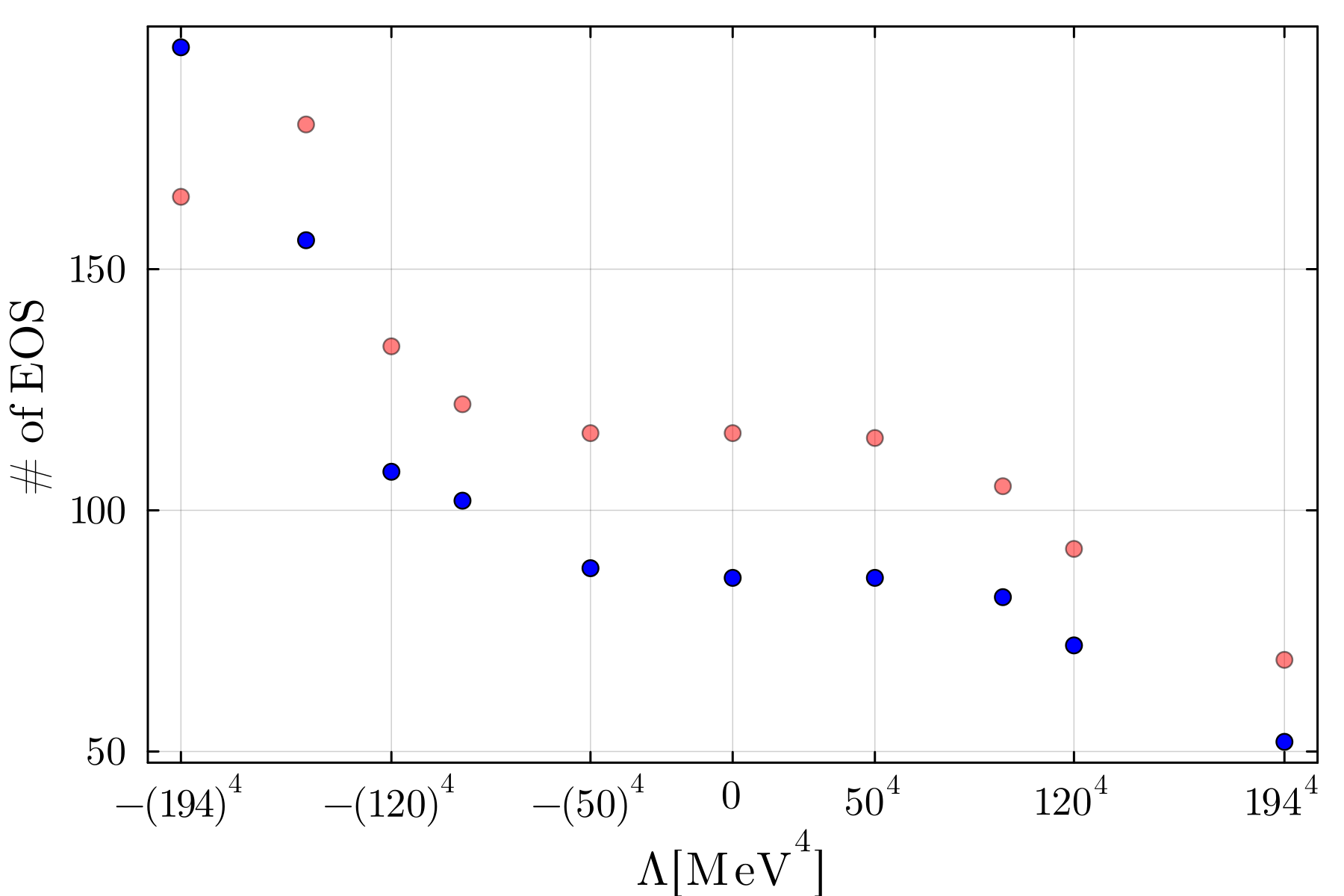}
\caption{The left (right) panel describes the smaller (larger) dataset. Red (blue) dots corresponds to EOSs built starting from the SLy (AP4). We see that for positive $\Lambda$ the number of EOSs decreases fast. For negative values of cosmological constant, the number either remains constant for $\Lambda \gtrsim-(120\, \text{MeV})^4$ or drastically increase for $\Lambda  \lesssim -(150\, \text{MeV})^4$.}
	\label{fig:100EOS}
\end{center}
\end{figure*}

In this section, we present a first analysis on the EOSs dataset. As we discussed in section~\ref{sec:Impl}, we perform a first selection on the EOSs based on the maximum mass they can yield. This in fact drastically reduces the number of EOSs for each dataset. For the smaller dataset, of the EOSs built starting from the AP4, only $9\%$ survived, while for those constructed from the SLy, $9.7\%$ passed the selection, for a total of a $9.3\%$ of the full $2\times 10^3$ EOSs dataset. For the larger dataset we find similar results, of the total EOSs only $11.21\%$ passed the mass test ($10.3\%$ for the AP4 and $12.1\%$ for the SLy). The reason for this is that the new part of the EOS describing the QCD phase transition built by random segments often forces the maximum mass to be considerably smaller than what the standard AP4 or SLy yields. Hence, this strict selection seems to be related to our choice for describing the new QCD matter phase, rather than a feature of implementing a vacuum energy shift. As a matter of fact, introducing a sufficiently negative cosmological constant contribution in the core can save these kind of EOSs, as we will show. This is because the EOS becomes stiffer for a negative cosmological constant, as it contributes positively to the pressure and so matter is less compact and the star can support larger masses. 

\begin{figure}[]
\centering
    \includegraphics[width=1.0\linewidth]{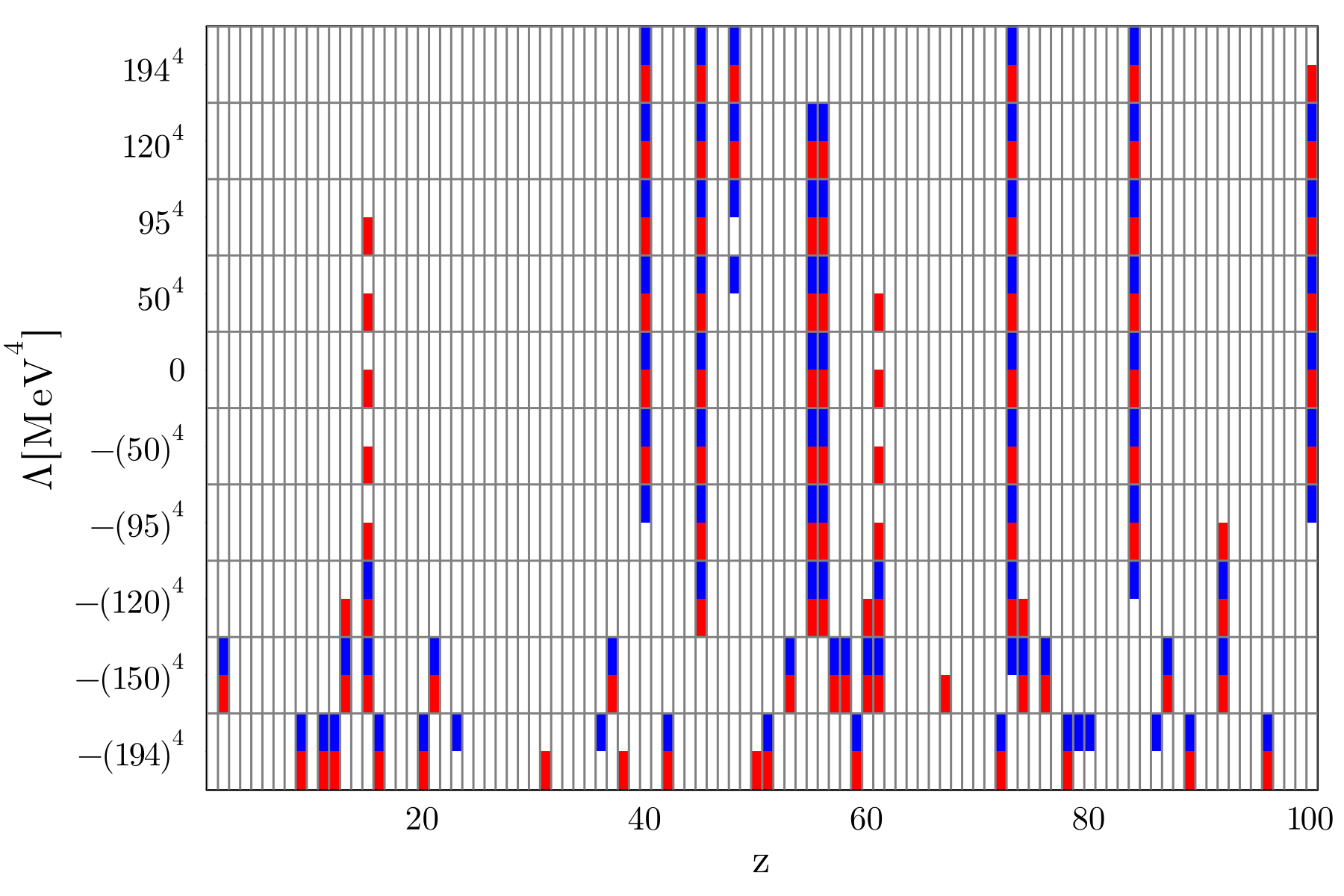}%
    \caption{Summary of all possible configurations for the smaller dataset. The label `z' is a bookkeeping parameter, corresponding to a specific choice of 6 random segments in the extension to high-densities. A red (blue) block correspond to an admissible EOS constructed from the SLy (AP4). We can identify two different EOSs families: EOSs that allow for a wide range of $\Lambda$, and EOSs that can exist only for very negative value of vacuum energy jump.}
	\label{fig:fulldataset}
\end{figure}

Let us now focus on the `admissible' EOSs. In Fig.~\ref{fig:100EOS} we show the number of EOSs that passes the mass test versus the value of $\Lambda$. The left (right) panel shows the results for the smaller (larger) dataset. For $\Lambda>0$, the number of `admissible' EOSs tends to either remain constant or decrease with increasing $\Lambda$ for both EOSs built starting from the AP4 and the SLy. Note that for large positive values of vacuum energy jump, e.g., $\Lambda = (194\, \text{MeV})^4$, only a very small fraction of EOS survives the maximum mass test for both dataset. As we will see in the next section, this is due to the fact that positive values of shift in vacuum energy lower the maximum mass, the larger $\Lambda$ the bigger the decrease.
For $\Lambda < 0$, the situation is more interesting. There are two distinct behaviour; for $\Lambda \gtrsim - (120\, \text{MeV})^4$ the number of admissible EOS tends to remain constant, whereas there is a sudden peak for very small value of $\Lambda$ , e.g., $\Lambda = -(194\, \text{MeV})^4$. We can better understand this behaviour by looking at Fig.~\ref{fig:fulldataset}, where we schematically report which configurations are admissible for the smaller dataset. Note that we have defined a label `z' as a bookkeeping parameter, corresponding to a specific choice of 6 random segments in the extension to high-densities. A red (blue) block corresponds to EOS constructed from the SLy (AP4). The vacuum energy jump in the core creates two families of EOSs: on one side we find EOSs that allow for a wide range of cosmological constant values, including $\Lambda=0$, on the other one we obtain EOSs that can exist only if a very negative jump in vacuum energy is introduced. We further investigate this mechanism in the next section.

\subsection{M-R curve}\label{sec:MR}

\begin{figure*}[t!]
\begin{center}	\includegraphics[width=0.5\linewidth]{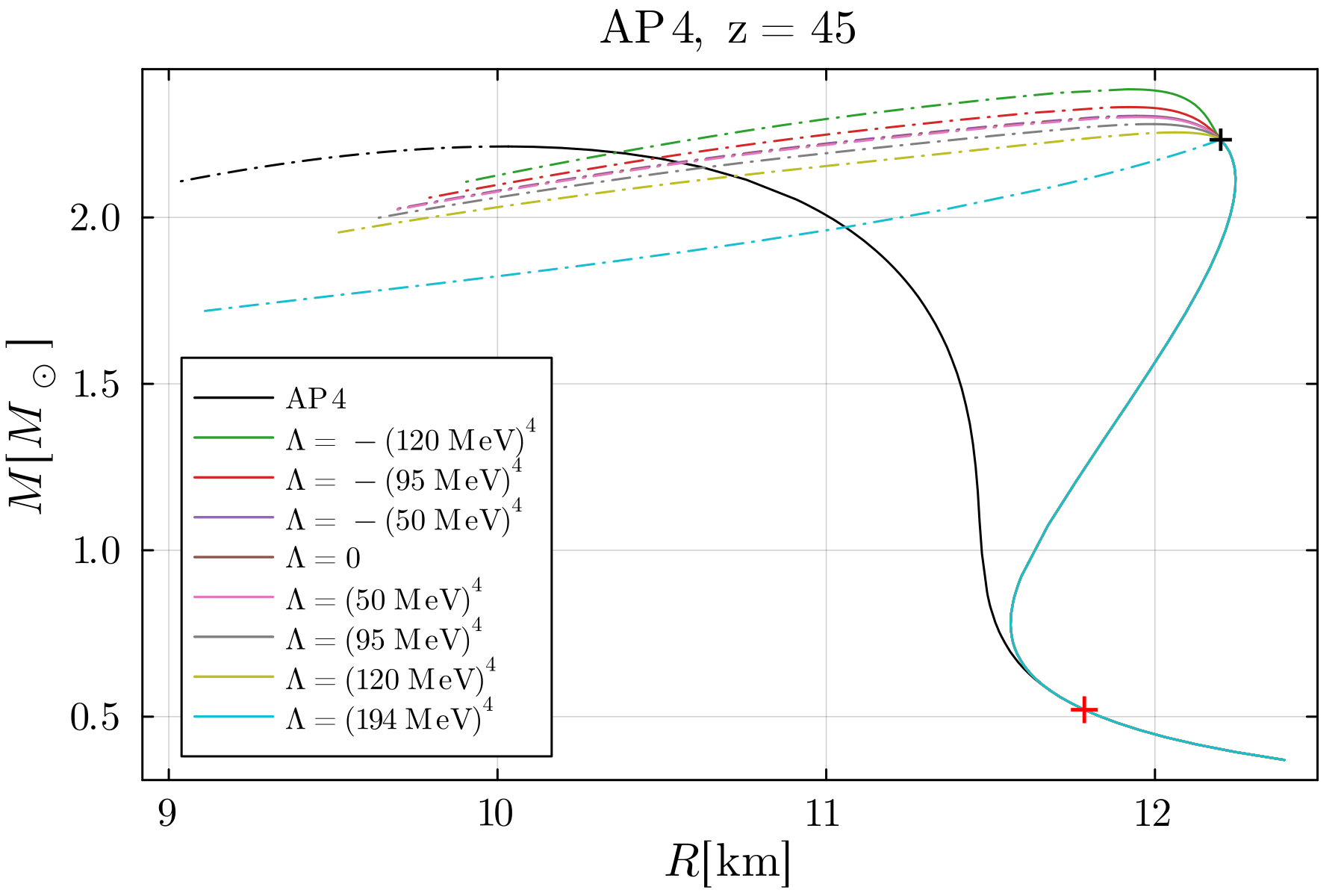}\hfill
\includegraphics[width=0.5\linewidth]{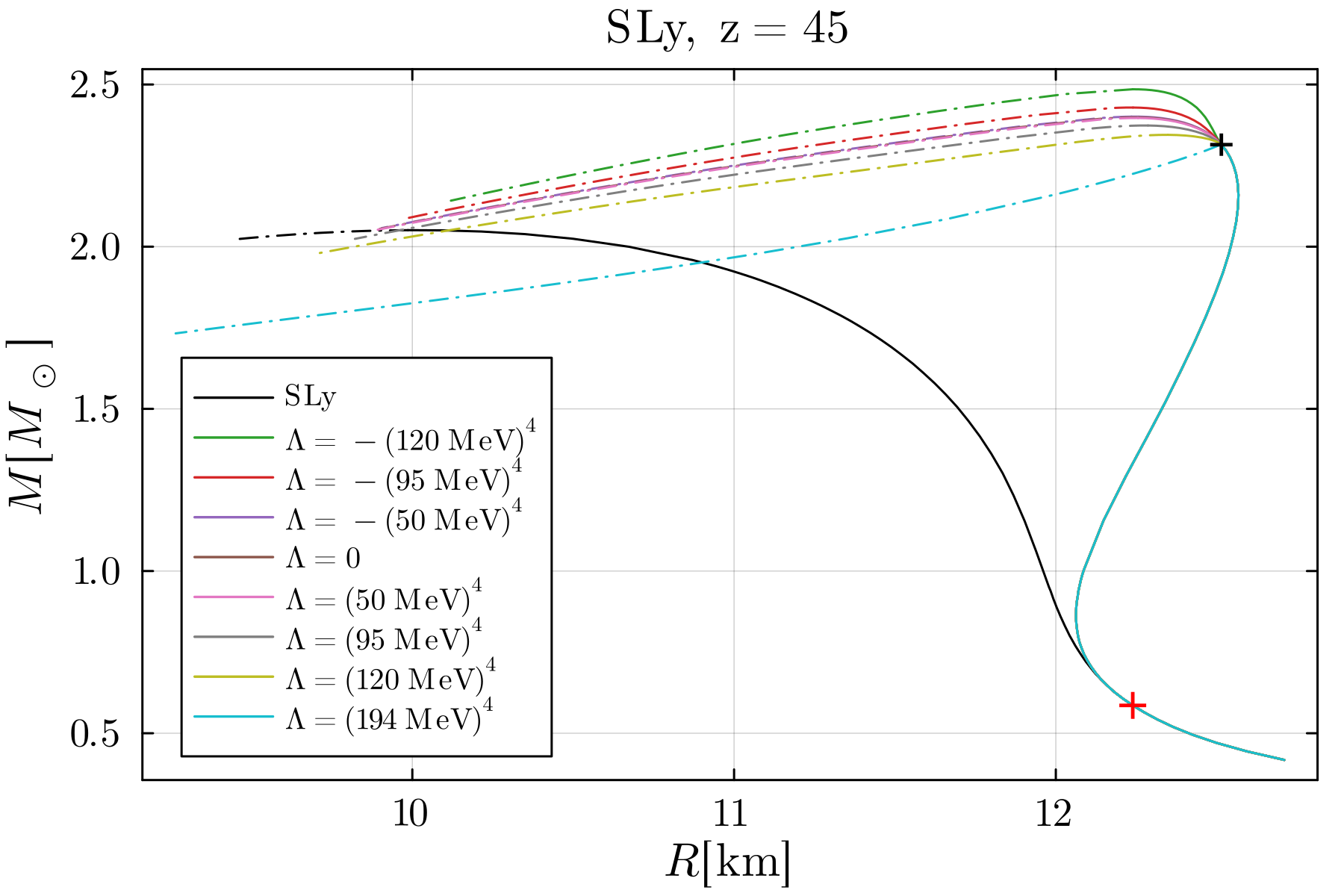}
	\caption{Mass-radius curves for static stars for a specific high-density EOS modifications for different value of vacuum energy. In the left (right) panel we show the results for the AP4 (SLy) as low-density description. A black line corresponds to the standard AP4 or SLy EOS. The two phase transitions, QCD and vacuum energy, bifurcation points are described by a red and black cross respectively. A dash-dotted line corresponds to unstable configurations.}
	\label{fig:MvsR45}
\end{center}
\end{figure*}

We now consider the effects of $\Lambda$ on the mass versus radius curve. We first focus on a specific baseline EOS, corresponding to label $z=45$ in Fig.~\ref{fig:fulldataset}.\footnote{We once again stress that the labels are simply bookkeeping parameters, corresponding to a specific choice of 6 random segments in the extension to higher densities.} Note that this baseline EOS belongs to the first family we discussed in section~\ref{sec:dataset}, i.e. it allows for a wide range of $\Lambda$. The results are summarized in Fig.~\ref{fig:MvsR45}, where we compare the curves of the modified EOSs with those obtained with standard AP4 and SLy (black lines). The figure shows a first transition from the standard EOSs to the high density modification and a second one from the jump in vacuum energy, corresponding to a red and a black cross respectively. A dash-dotted line corresponds to unstable configurations that violate the condition $\partial M/\partial p_0 > 0$, where $p_0$ is the central pressure. We first note that all the new EOSs are more stiff than the standard ones, i.e. given the same mass they allow for bigger radii. Furthermore, a negative (positive) $\Lambda$ in the core increases (decreases) the maximum mass that neutron stars can reach. This feature is a first difference to the results obtained in Ref.~\cite{Csaki:2018fls}, where all values of $\Lambda$ yielded a decrease in maximum mass. This feature in Ref.~\cite{Csaki:2018fls} is a consequence of the fixed positive jump in energy density when the transition kicks in. By allowing a negative $\Delta\epsilon$, we also allow neutron stars to sustain higher maximum masses. Note that when the $\Lambda=(194\,\text{MeV})^4$ transition is triggered the M-R curve immediately becomes unstable. We do not show it here, but this is true for all the EOSs we generated (built either from the SLy or the AP4) that admit such a large value of vacuum energy shift.

\begin{figure}[h]
\begin{center}
	\subfloat{%
	\includegraphics[width=1\linewidth]{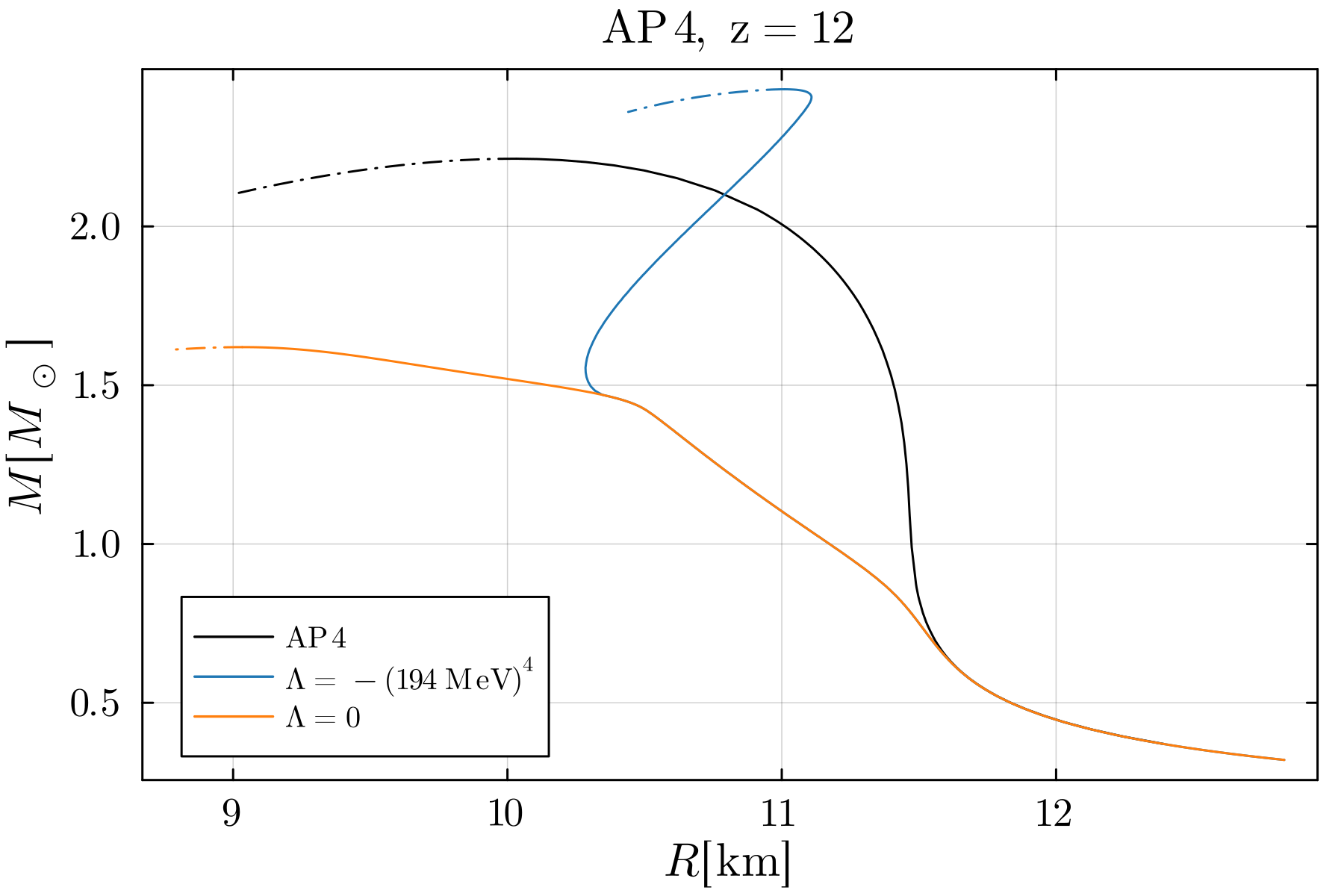}%
	}
  \caption{Mass-radius curve for static stars for a fixed high energy EOS modifications, with low-density described by the AP4, that allows for $\Lambda = -(194\, \text{MeV})^4$. A black line corresponds to the standard AP4.  The only allowed configurations, with $\Lambda = -(194\, \text{MeV})^4$, corresponds to a blue line. We added, for comparison, a configuration with $\Lambda = 0$, in orange; such choice of $\Lambda$ is not able to sustain a maximum mass in agreement with the constraints from PSR J0952-0607. A dash-dotted line corresponds to unstable configurations.}
	\label{fig:MvsRLfix}
\end{center}
\end{figure}

Let us now focus on the second family of EOSs, those that are allowed only for large negative values of $\Lambda$. We show the results in Fig.~\ref{fig:MvsRLfix} for a specific modified EOS with the AP4 describing low-density matter, corresponding to label $z=12$ in Fig.~\ref{fig:fulldataset}. Other choices of EOSs lead to agreeing conclusions. The figure compares the mass-radius curve for $\Lambda=-(194\, \text{MeV})^4$ with the standard AP4 and the baseline EOS, corresponding to $\Lambda=0$. The latter yields a maximum mass which is too low to be in agreement with current observations. However, implementing a large negative jump in vacuum energy in the core produce a sufficiently high maximum mass, thus `saving' the configuration.

\begin{figure*}[t!]
\begin{center}	\includegraphics[width=0.5\linewidth]{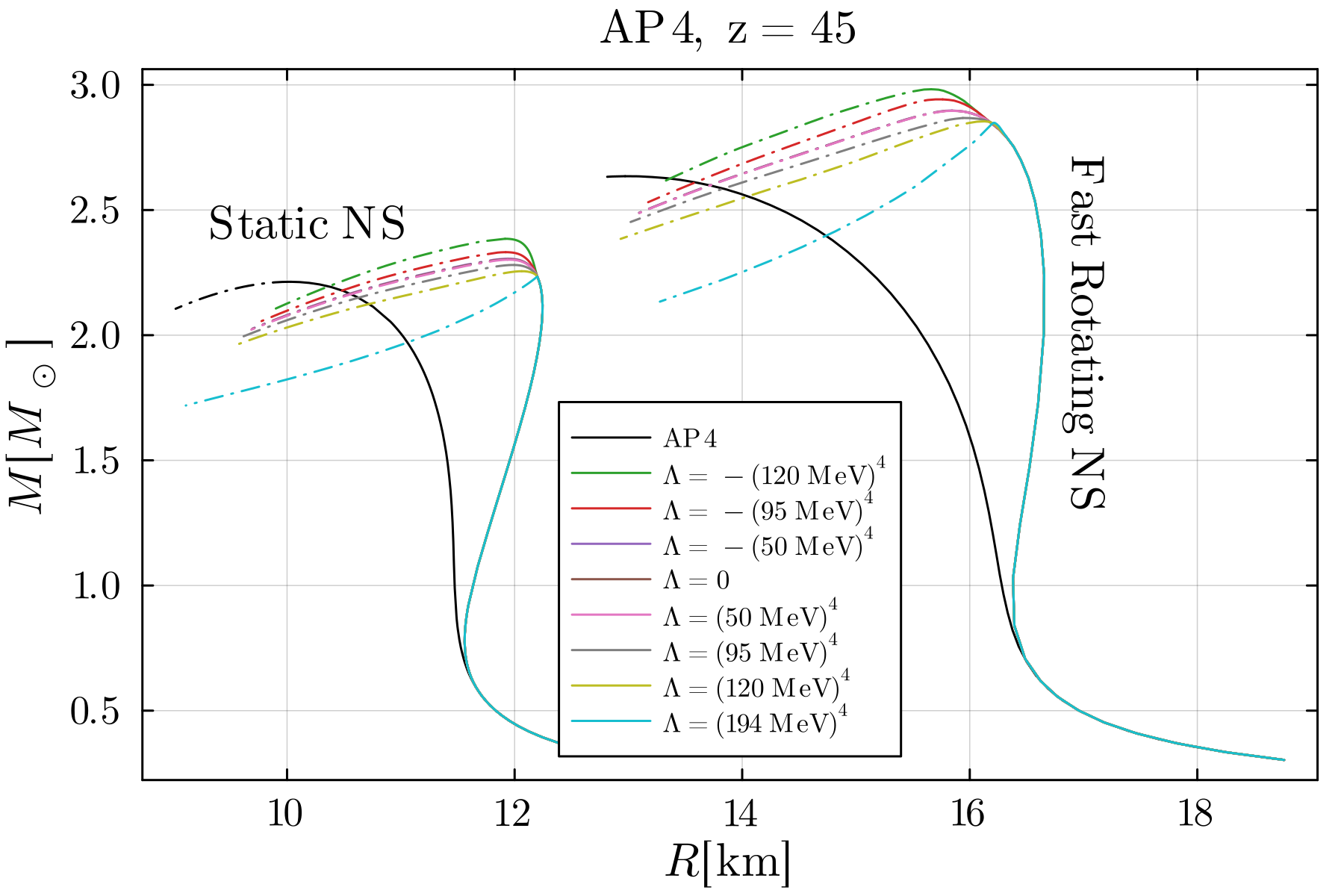}\hfill
\includegraphics[width=0.5\linewidth]{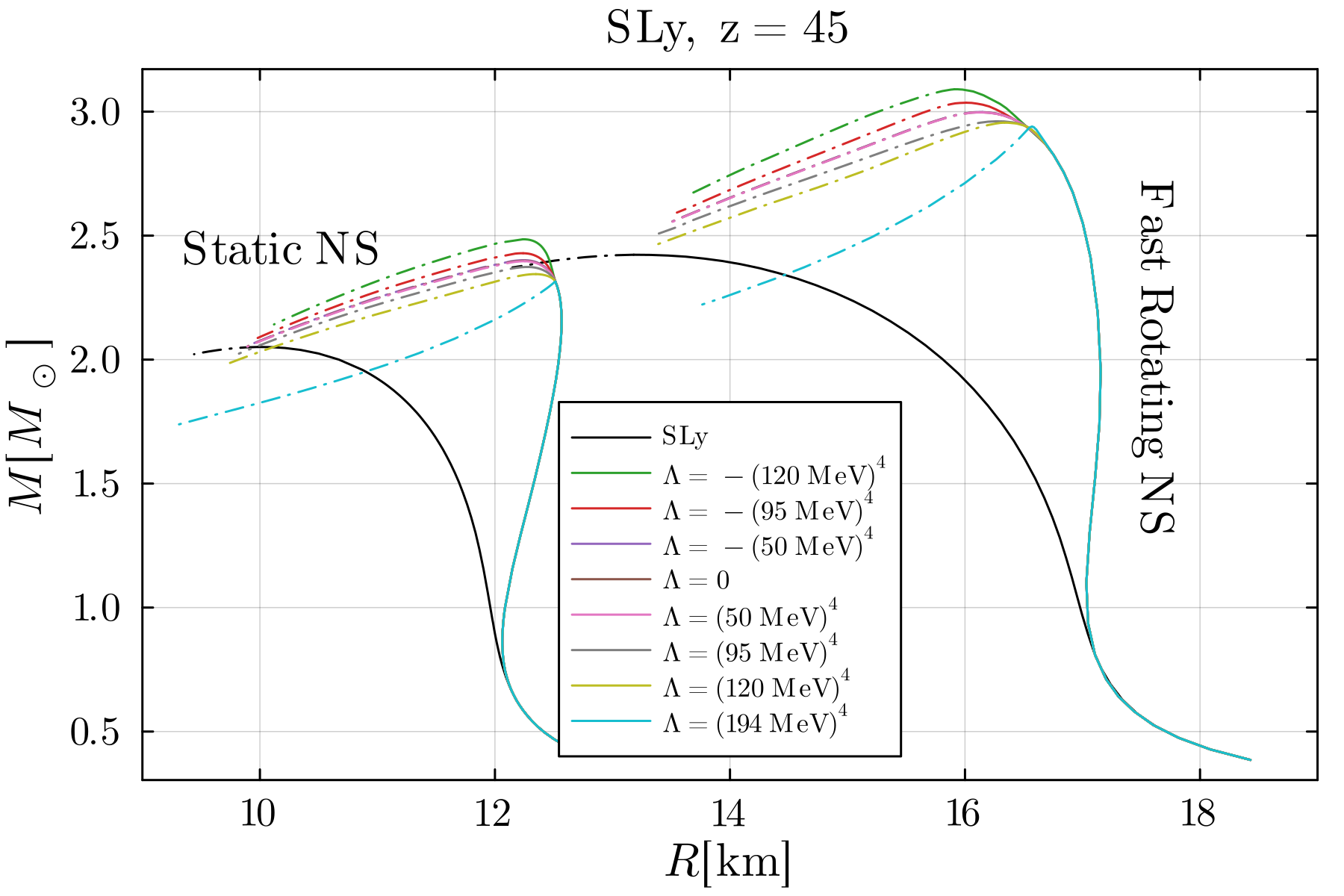}
	\caption{Mass-radius curves for both static and maximally rotating neutron stars for a specific high-density EOS modification for different values of vacuum energy. In the left (right) panel we show the results for the AP4 (SLy) as low-density description. A black line corresponds to the standard AP4 or SLy EOS. A dash-dotted line corresponds to unstable configurations.}
	\label{fig:MvsR45_Rot}
\end{center}
\end{figure*}

The behaviour observed for rotating neutron stars is qualitatively similar, as it is shown in Fig.~\ref{fig:MvsR45_Rot} for the specific baseline EOS corresponding to the label $z=45$. The main difference observed in the mass-radius curves between static and rotating stars is that the curves for rotating stars are shifted to the right (i.e., for the same mass they have larger radii), and allow for more massive configurations. Concerning the different vacuum energy values in the core, the conclusions of the static case hold; all new EOSs are more stiff than the standard ones and a negative (positive) $\Lambda$ in the core increases (decreases) the maximum mass that a neutron star can reach.

\subsection{Tidal deformability}\label{sec:tidalDef}

\begin{figure}
    \centering
    \includegraphics[width=1\linewidth]{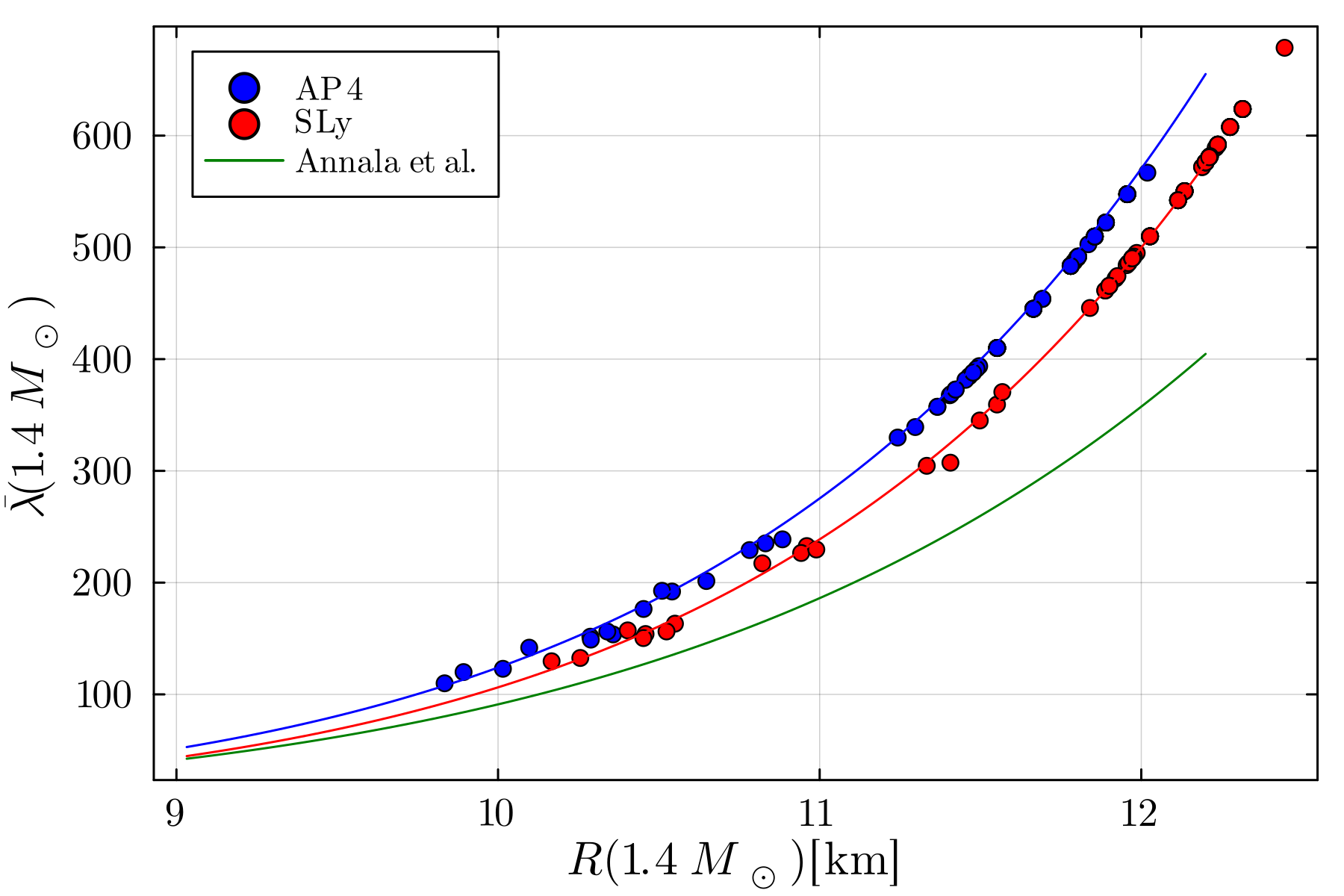}
    \caption{$\bar{\lambda}$ vs $R$ for stars with $M=1.4 M_{\odot}$ for all admissible configurations in the smaller dataset. For this value of mass, the vacuum energy contributions do not come into play. Blue (red) dots correspond to EOSs built from the AP4 (SLy). We also show the respective fits. The green line is the empirical function obtained by Annala \textit{et al.}~\cite{Annala:2017llu}.}
    \label{fig:Annala}
\end{figure}

Let us now consider the effects on the tidal deformability. We first focus on the correlation between the dimensionless tidal deformability $\bar{\lambda}$ and radius $R$. Let us start by discussing the case where only the QCD transition takes place. Hence, we focus on neutron star configurations with the fixed mass $M = 1.4 M_\odot$, for which none of the modified EOSs have already developed the vacuum energy phase transition. The results are displayed in Fig.~\ref{fig:Annala}. The figure shows the tidal deformability $\bar{\lambda}$ as a function of the radius of the star $R$ for each configuration on the smaller dataset. We see a tight correlation between $\bar{\lambda}$ and $R$ for $M = 1.4 M_\odot$. For the AP4 ensemble all the tidal deformabilities follow the empirical function $\bar{\lambda}(R) = 5.26 \times 10^{-7} (R/\text{km})^{8.37}$, whereas for the SLy they follow $\bar{\lambda}(R) = 3.44 \times 10^{-7} (R/\text{km})^{8.49}$. Note that these fits are slightly different from previous results~\cite{Annala:2017llu}, where it was found that $\bar{\lambda}$ follows the empirical function $\bar{\lambda}(R) = 2.88 \times 10^{-6} (R/\text{km})^{7.5}$, which we report as a green line in Fig.~\ref{fig:Annala}. We suspect this is due to a different description for the low-density part of the EOS. The results in Fig.~\ref{fig:Annala} are in agreement with current gravitational waves observations from GW170817~\cite{LIGOScientific:2017vwq}. In Ref.~\cite{LIGOScientific:2017vwq}, the function $\bar{\lambda}(M)$ was constrained by linearly expanding it around $M=1.4\,M_\odot$ (as in Refs.~\cite{DelPozzo:2013ala,Agathos:2015uaa}), which gives $\bar{\lambda}(1.4 M_\odot) \lesssim 800$ for a low-spin prior. We do not report it here, but this also holds for the larger dataset.

\begin{figure}[h]
\begin{center}
	\subfloat{%
	\includegraphics[width=1\linewidth]{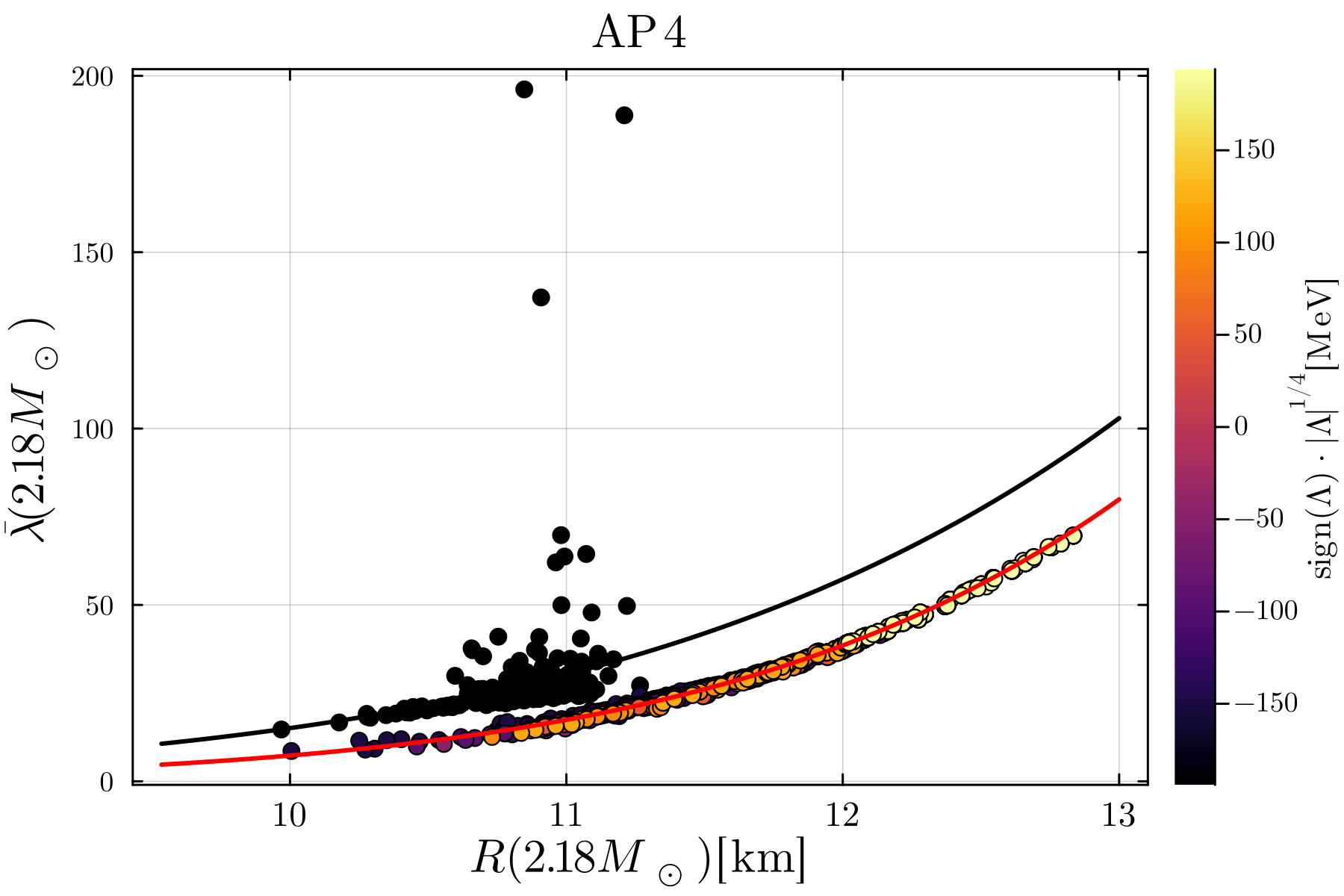}%
	}\\
	\subfloat{%
	\includegraphics[width=1\linewidth]{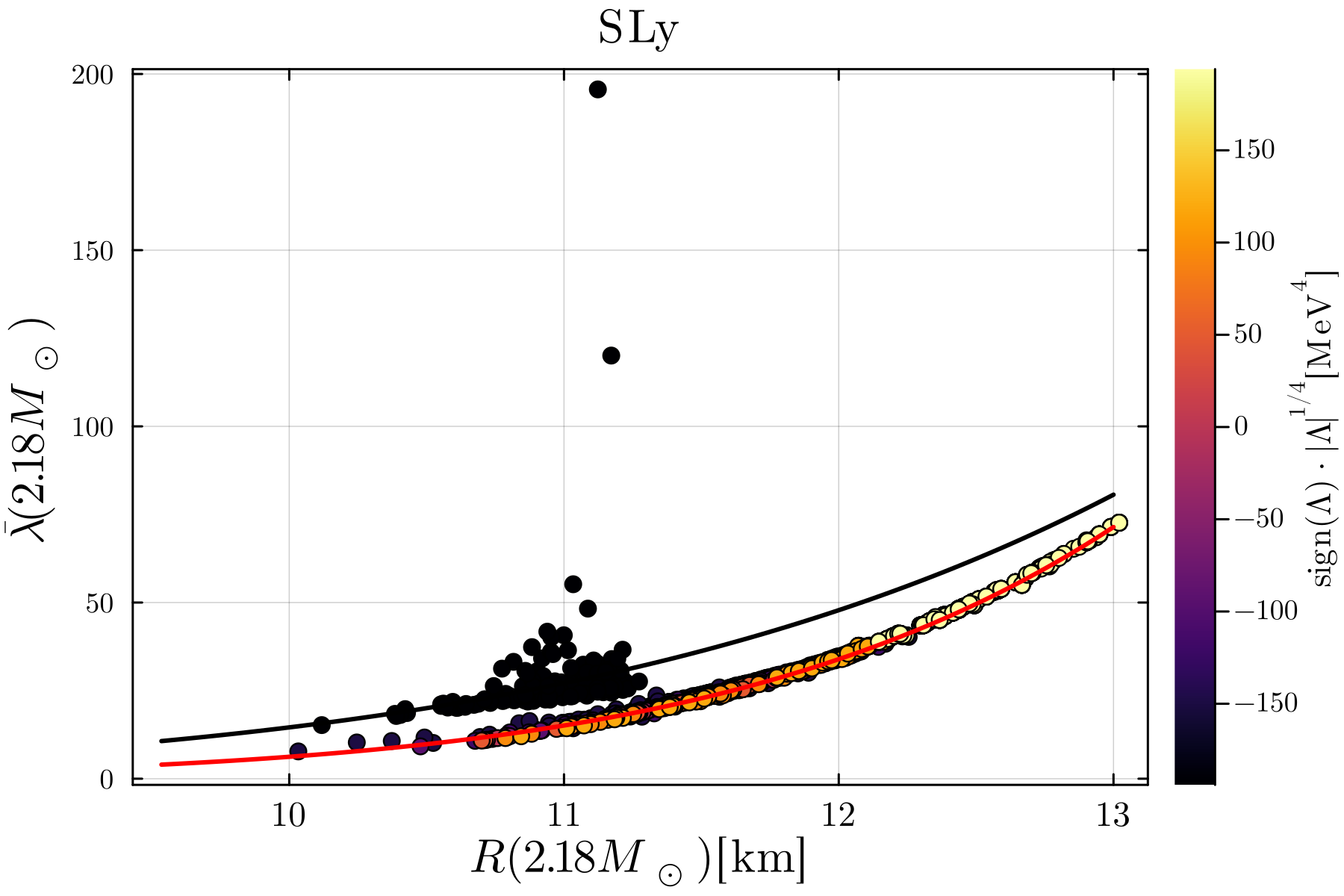}%
	}
	\caption{$\bar{\lambda}$ vs $R$ for stars with $M=2.18M_{\odot}$ for all admissible configurations on the larger dataset. Top (bottom) panel corresponds to EOSs built from the AP4 (SLy). The $\Lambda$ dependence is highlighted by a colour map.}
	\label{fig:lambdavsR1000}
\end{center}
\end{figure}

To study neutron stars configurations affected by a vacuum energy phase transition, we focus on stars with higher central pressure, reaching higher masses. We then consider neutron stars with a fixed mass of $M = 2.18 M_\odot$. We show the results for the larger dataset in Fig.~\ref{fig:lambdavsR1000}. In the top (bottom) panel we show the results for EOSs with the low-density region described by the AP4 (SLy), while also highlighting the dependence on the vacuum energy shift $\Lambda$ with the use of a colour map. In both cases there seem to be two distinct behaviours.

Let us focus on the AP4 case. The figure shows that the lower (red) curve, described by the empirical function $\bar{\lambda}(R) = 5.54 \times 10^{-9} (R/\text{km})^{9.12}$, is followed by neutron stars configurations independently of the values of $\Lambda$, apart from the case where $\Lambda = -(194\, \text{MeV})^4$. In the latter instance, the data follow a different pattern. Employing a power-law fit, we find that the majority of the data follow the empirical function (in black) described by $\bar{\lambda}(R) = 7.22 \times 10^{-7} (R/\text{km})^{7.32}$, with  some anomalous data points yielding much larger values of $\bar \lambda$. This confirms how these two are different families of EOSs, and they seem to develop different properties. For the case of the SLy the two curves are respectively $\bar{\lambda}(R) = 3.15 \times 10^{-9} (R/\text{km})^{9.30}$ and $\bar{\lambda}(R) = 4.40 \times 10^{-6} (R/\text{km})^{6.52}$.

\begin{figure*}[t!]
\begin{center}	
\includegraphics[width=0.5\linewidth]{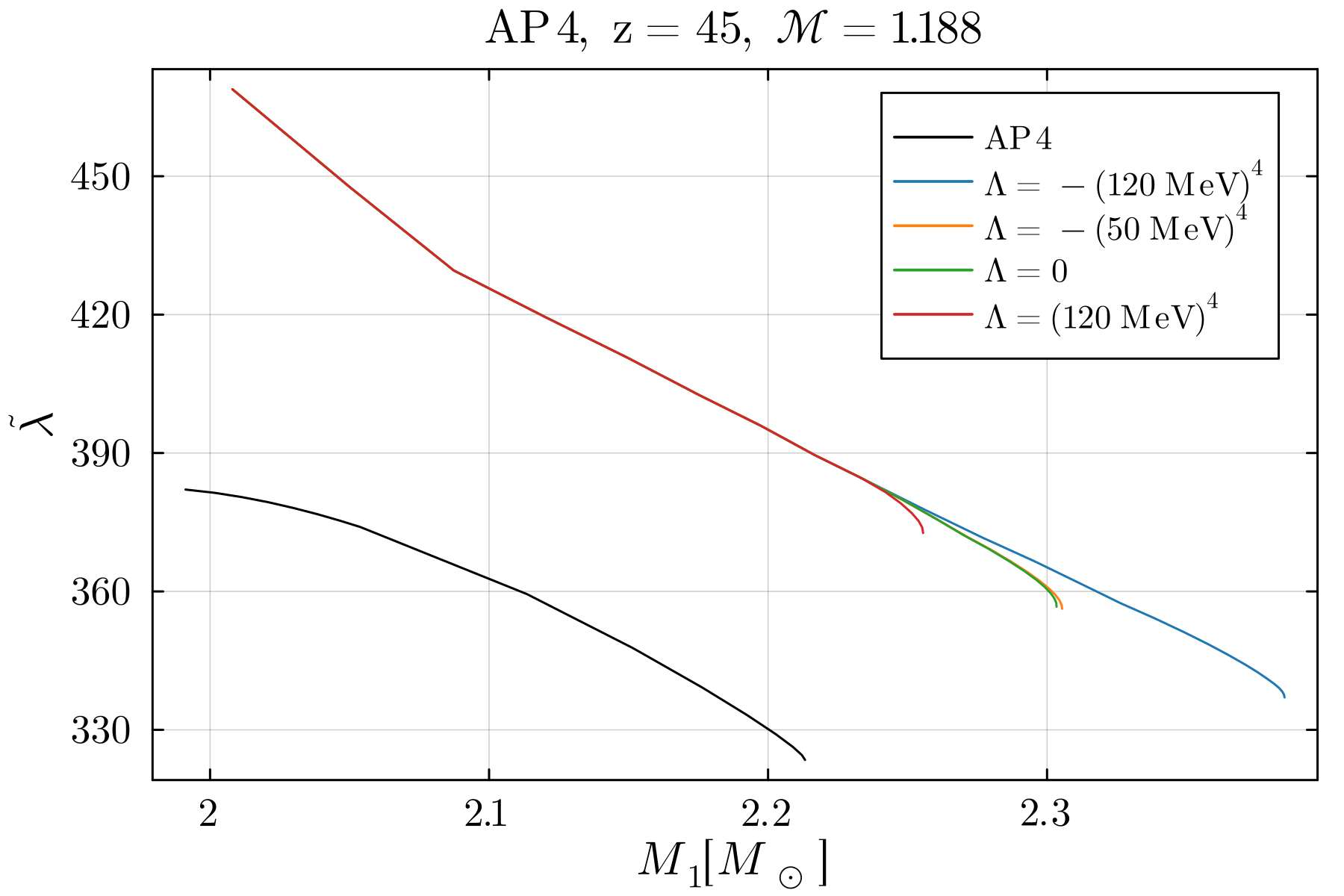}\hfill
\includegraphics[width=0.5\linewidth]{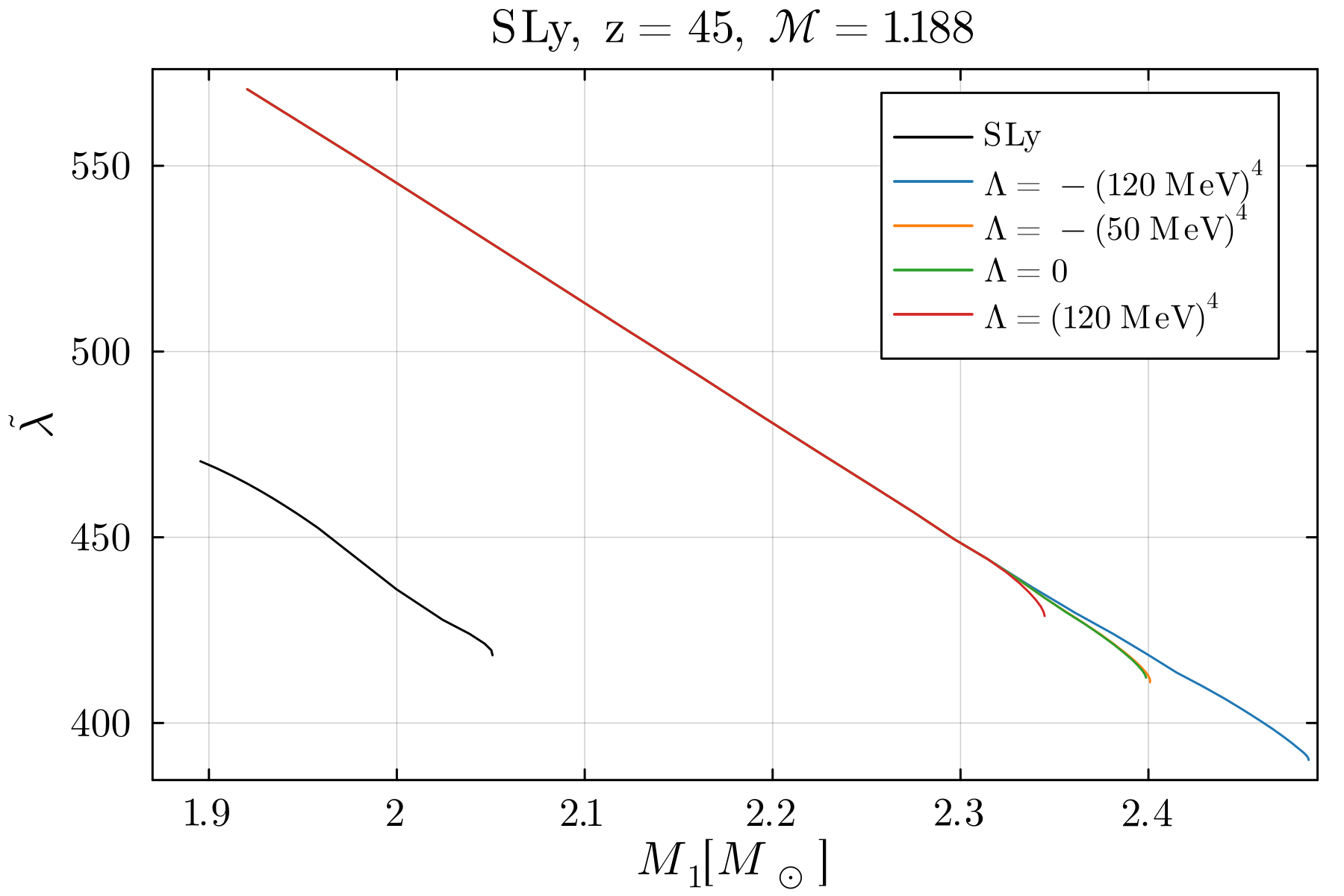}
	\caption{Deviation of combined dimensionless tidal deformability as a function of the mass of the primary in the binary, for a specific high-density EOS modification. In the left (right) panel we show the results for the AP4 (SLy) as low-density description. We consider different values of vacuum energy shift $\Lambda$, which we compare to the vanilla EOSs (black lines). The choice of chirp mass is $\mathcal{M}=1.188$, corresponding to that of GW170817.}
	\label{fig:CombTid_1.188}
\end{center}
\end{figure*}

\begin{figure*}[t!]
\begin{center}	\includegraphics[width=0.5\linewidth]{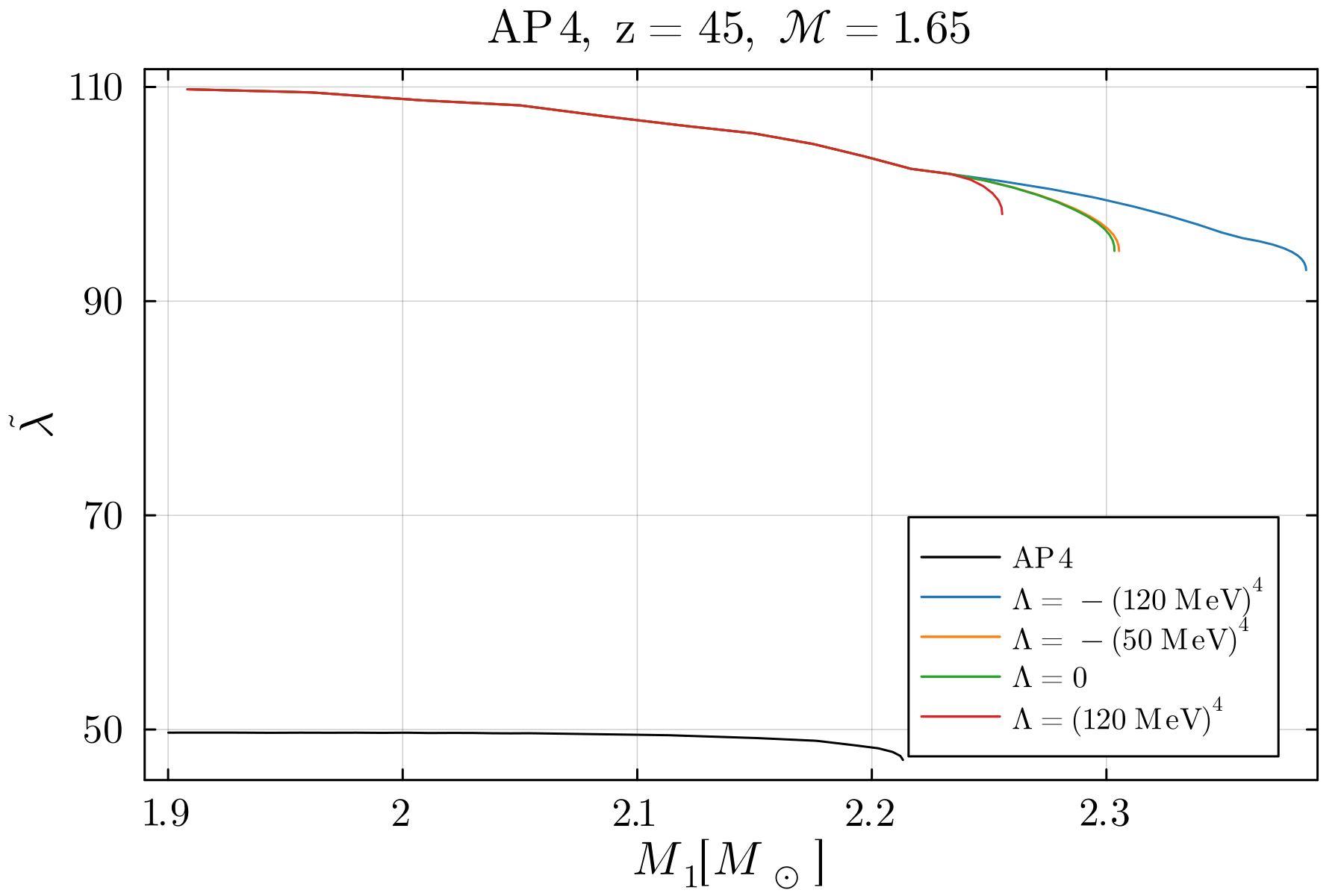}\hfill
\includegraphics[width=0.5\linewidth]{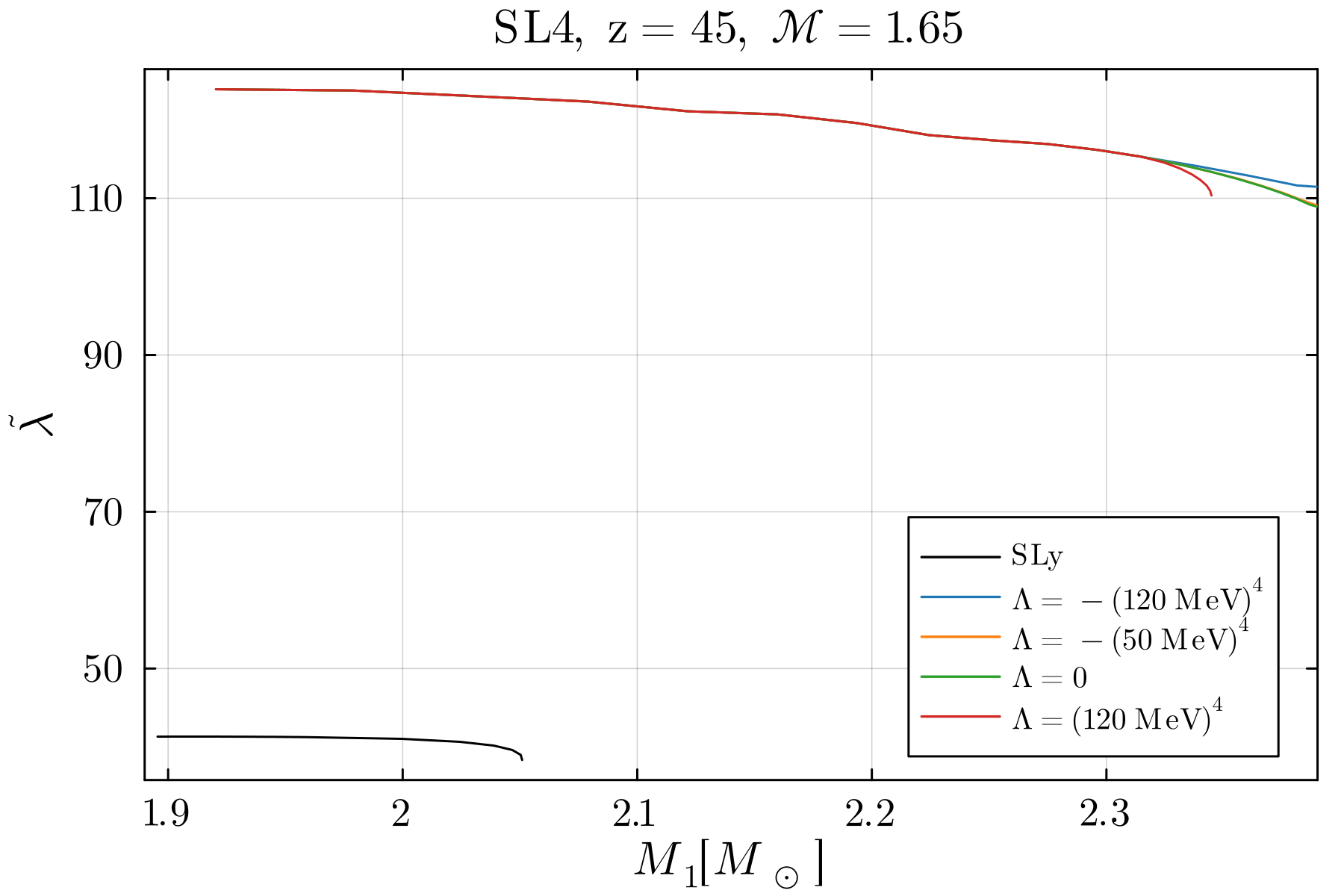}
	\caption{Deviation of combined dimensionless tidal deformability as a function of the mass of the primary in the binary, for a specific high-density EOS modification. In the left (right) panel we show the results for the AP4 (SLy) as low-density description. We consider different values of vacuum energy shift $\Lambda$, which we compare to the vanilla EOSs (black lines). The choice of chirp mass is $\mathcal{M}=1.65$.}
	\label{fig:CombTid_1.65}
\end{center}
\end{figure*}

We now focus on observables from gravitational waves observation during the merger of a neutron stars binary. To leading order in the dimensionless tidal deformabilities of each companion, the gravitational wave phase is determined by the so-called ``combined dimensionless tidal deformability'', defined as~\cite{LIGOScientific:2017vwq,LIGOScientific:2018hze,Flanagan:2007ix,Favata:2013rwa}
\begin{multline}
\tilde{\lambda} = \frac{16}{13} \big[(M_1+12\,M_2)M_1^4\bar{\lambda}_1\\
+ (M_2 + 12\,M_1)M_2^4\bar{\lambda}_2\big]/(M_1 + M_2)^5,
\end{multline}
where a subscript 1 (2) corresponds to the primary (secondary) component of the binary. From the event GW170817, the current constraint on the combined dimensionless tidal deformability is $\tilde{\lambda}=300^{+500}_{-190}$, assuming a low-spin prior~\cite{LIGOScientific:2018hze}. In Fig.~\ref{fig:CombTid_1.188}, we show the dependence of $\tilde{\lambda}$ on the mass of the primary $M_1$ for a specific baseline EOS, corresponding to label $z=45$ in Fig.~\ref{fig:fulldataset}, constructed from both the AP4 (left panel) and the SLy (right panel) as low-density EOSs and for different choices of vacuum energy shift $\Lambda$. We choose a chirp mass $\mathcal{M} = \frac{(M_1\,M_2)^{3/5}}{(M_1+M_2)^{1/5}}$ corresponding to that of the event GW170817, that is $\mathcal{M}=1.188$. We compare the result with the standard AP4 and SLy (black curve). We first notice that already the QCD transition in the EOS introduces large deviations with respect to the standard EOS (AP4 or SLy). Once the vacuum energy shift is also triggered, large deviations from the baseline EOS appear for large values of $\Lambda$. Note that, for all cases, the results are in agreement with the current bound set from the event GW170817 ($\tilde{\lambda}=300^{+500}_{-190}$). The same conclusions hold when we consider a higher chirp mass. We show this in Fig.~\ref{fig:CombTid_1.65}, where we consider the same choice of EOSs for a chirp mass of $\mathcal{M}=1.65$, similarly to what was done in Ref.~\cite{Csaki:2018fls}. In Appendix~\ref{Appendix:RelativeShift} we show the relative shift of $\tilde{\lambda}$ for the case of EOSs built from the AP4.

\subsection{I-Love-Q relations}

In Refs.~\cite{Yagi:2013bca,Yagi:2013awa} it was first pointed out that suitable dimensionless combinations of the moment of inertia $I$, the tidal Love number $\lambda$ and the spin-induced quadrupole moment $Q$ satisfy universal relations, the so-called I-Love-Q relations, i.e. they do not depend on the EOS with an accuracy of a few percent~\cite{Lattimer:2012xj} (other approximate universal relations have been found between differrent stellar observables, see~\cite{Yagi:2016bkt} for a review). Here, we investigate if these relations still hold when introducing a new phase transition to the EOS.

\begin{figure*}[h!]
\begin{center}	\includegraphics[width=0.5\linewidth]{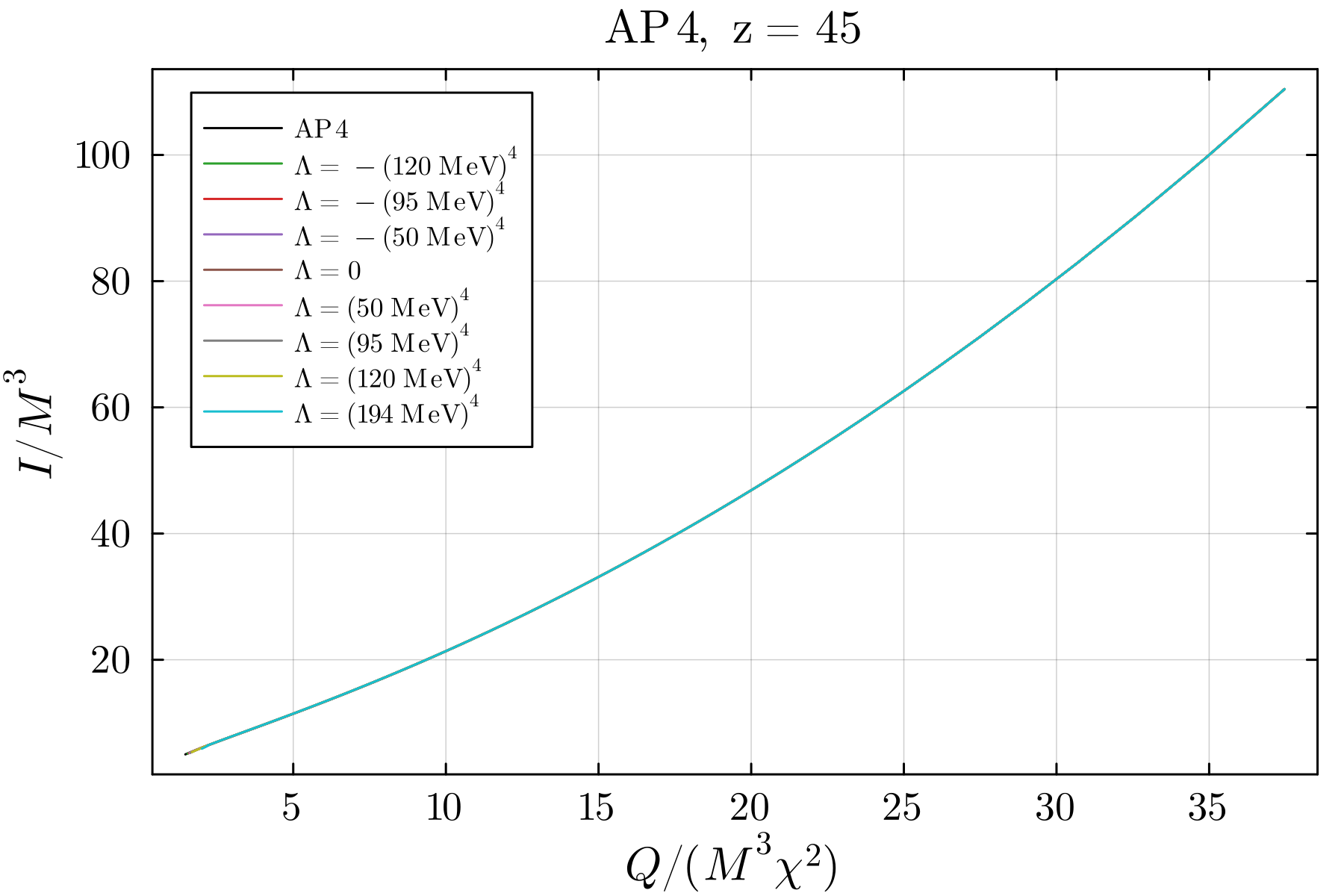}\hfill
\includegraphics[width=0.5\linewidth]{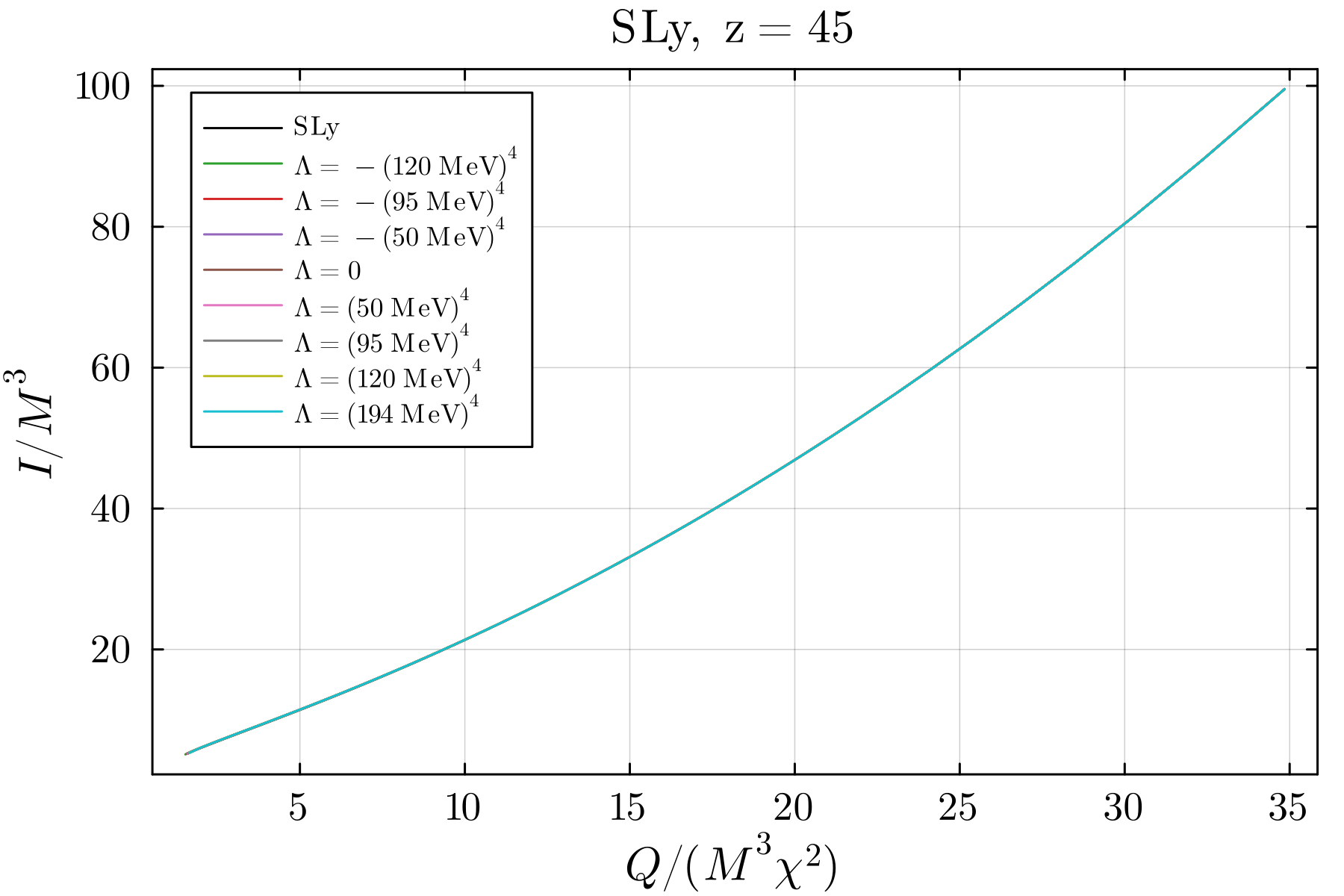}
	\caption{$I$ vs $Q$ for slowly rotating stars for a specific high-density EOS modification. The left (right) panel corresponds to the AP4 (SLy) as low-density description. We show the results for different values of vacuum energy shift, where differences are barely perceptible.}
	\label{fig:IQ45_SlowRot}
\end{center}
\end{figure*}

\begin{figure*}[h!]
\begin{center}	\includegraphics[width=0.5\linewidth]{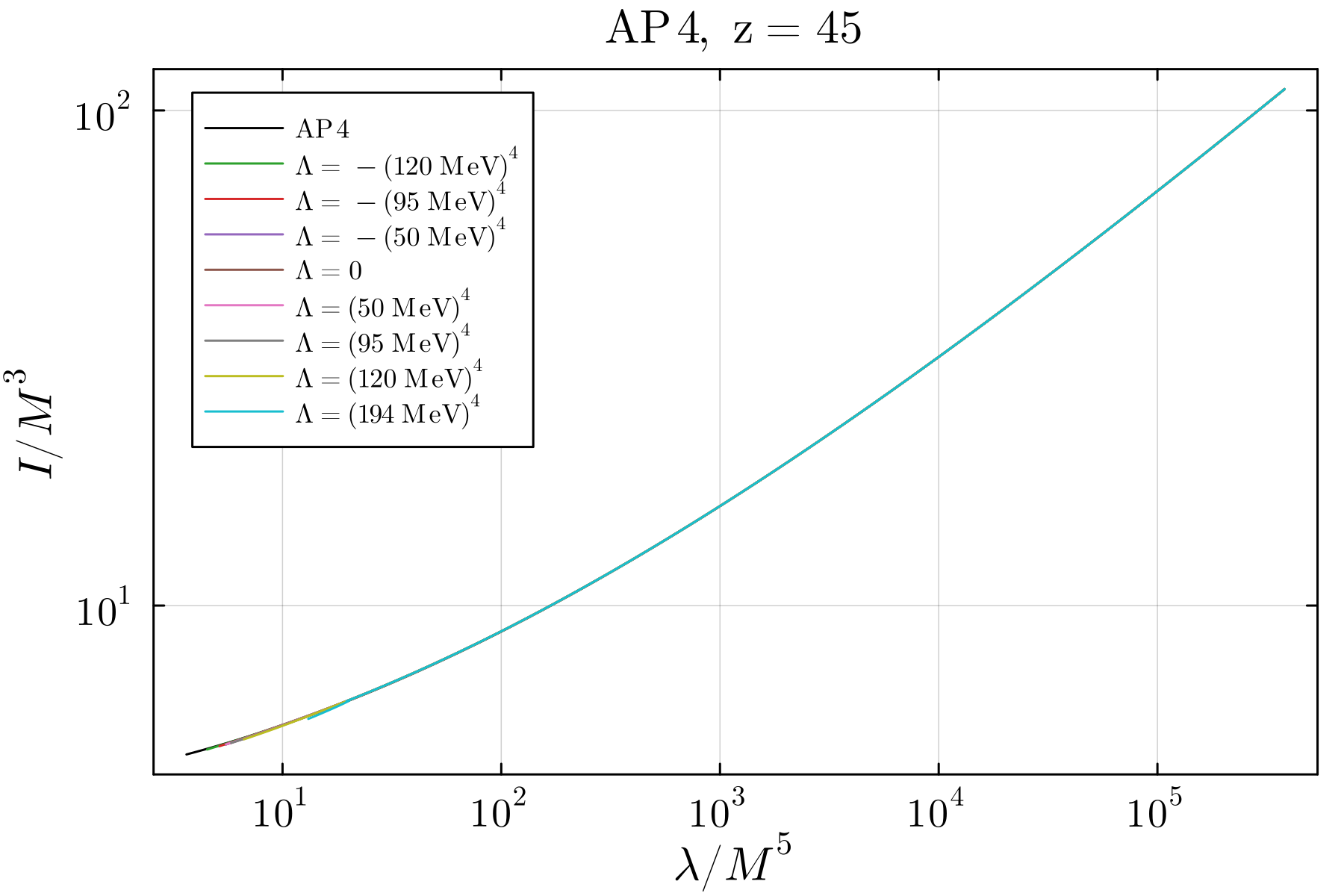}\hfill
\includegraphics[width=0.5\linewidth]{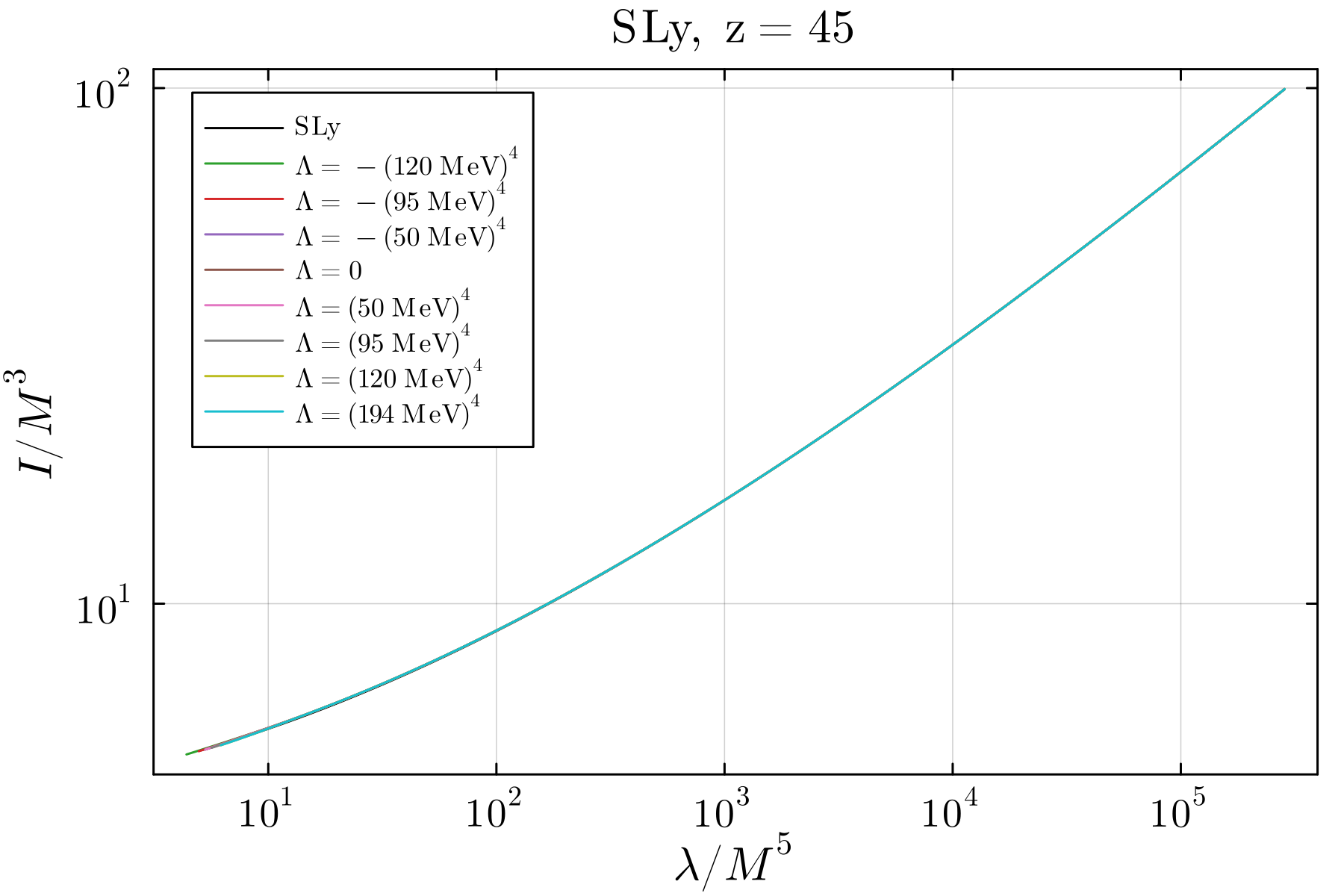}
	\caption{$I$ vs $\lambda$ for slowly rotating stars for a specific high-density EOS modification for different values of vacuum energy shift. The left (right) panel corresponds to the AP4 (SLy) as low-density description.}
	\label{fig:IL45_SlowRot}
\end{center}
\end{figure*}

\begin{figure*}[h!]
\begin{center}	\includegraphics[width=0.5\linewidth]{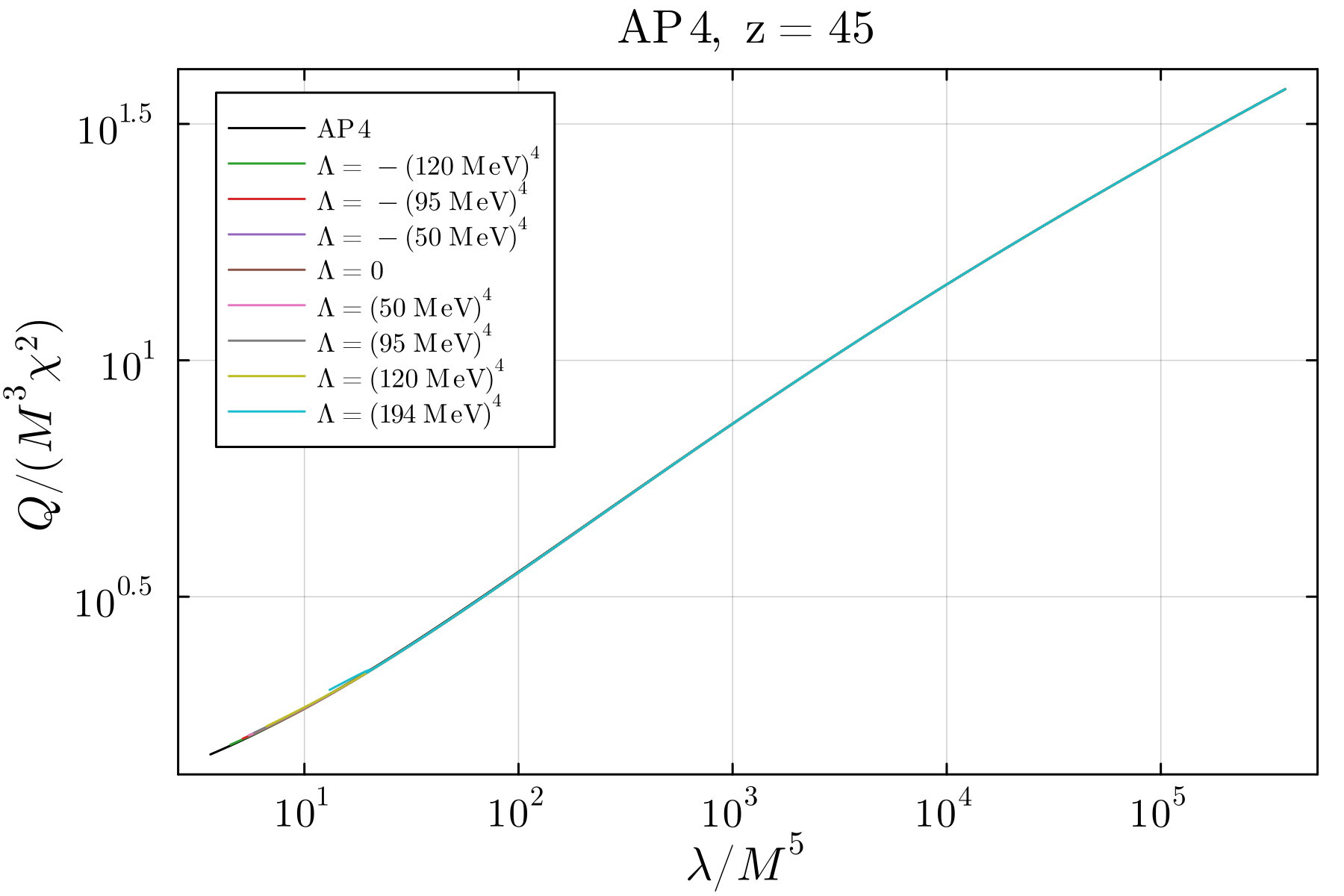}\hfill
\includegraphics[width=0.5\linewidth]{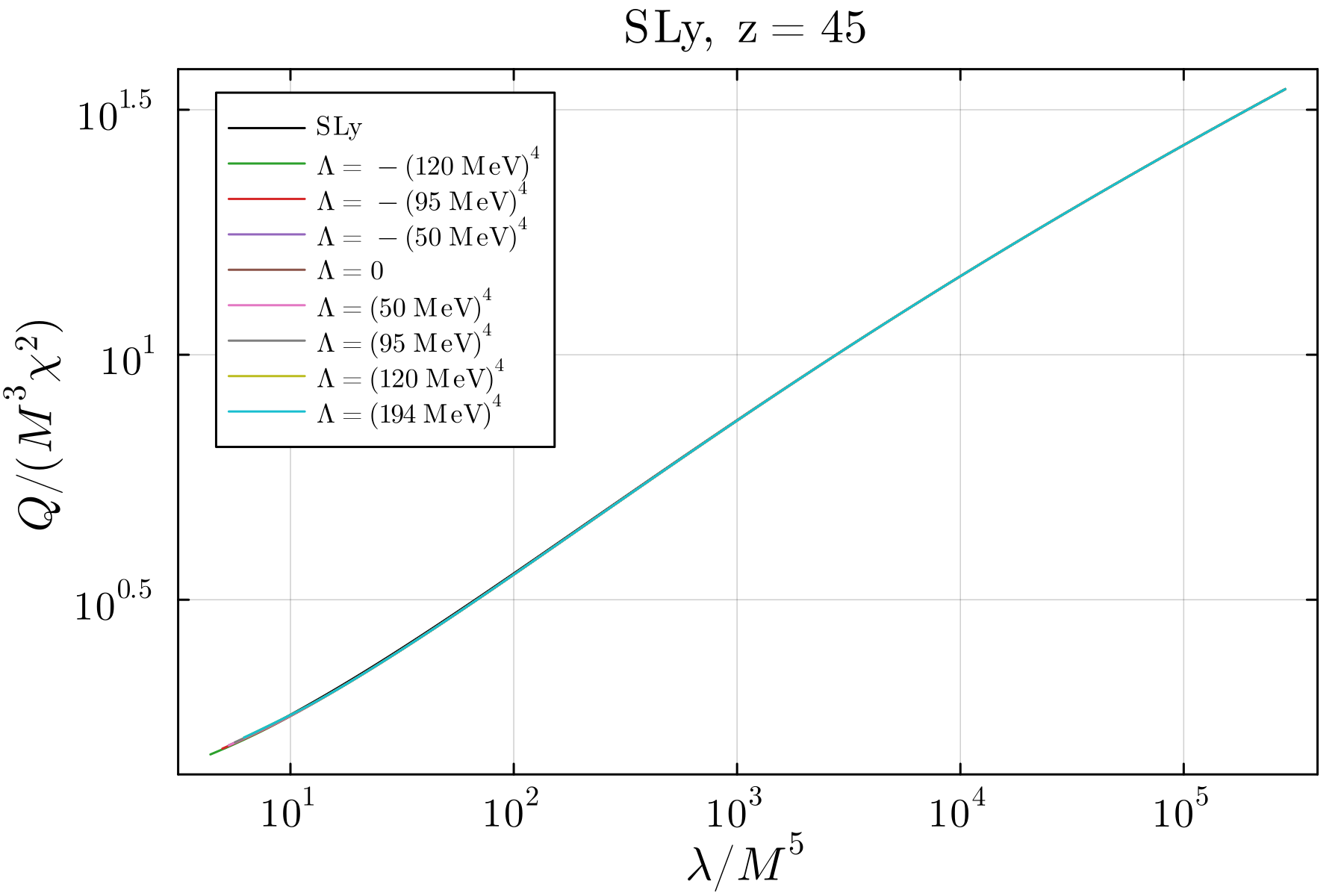}
	\caption{$Q$ vs $\lambda$ for slowly rotating stars for a specific high-density EOS modification. The left (right) panel corresponds to the AP4 (SLy) as low-density description.}
	\label{fig:QL45_SlowRot}
\end{center}
\end{figure*}

\begin{figure*}[h!]
\begin{center}	\includegraphics[width=0.5\linewidth]{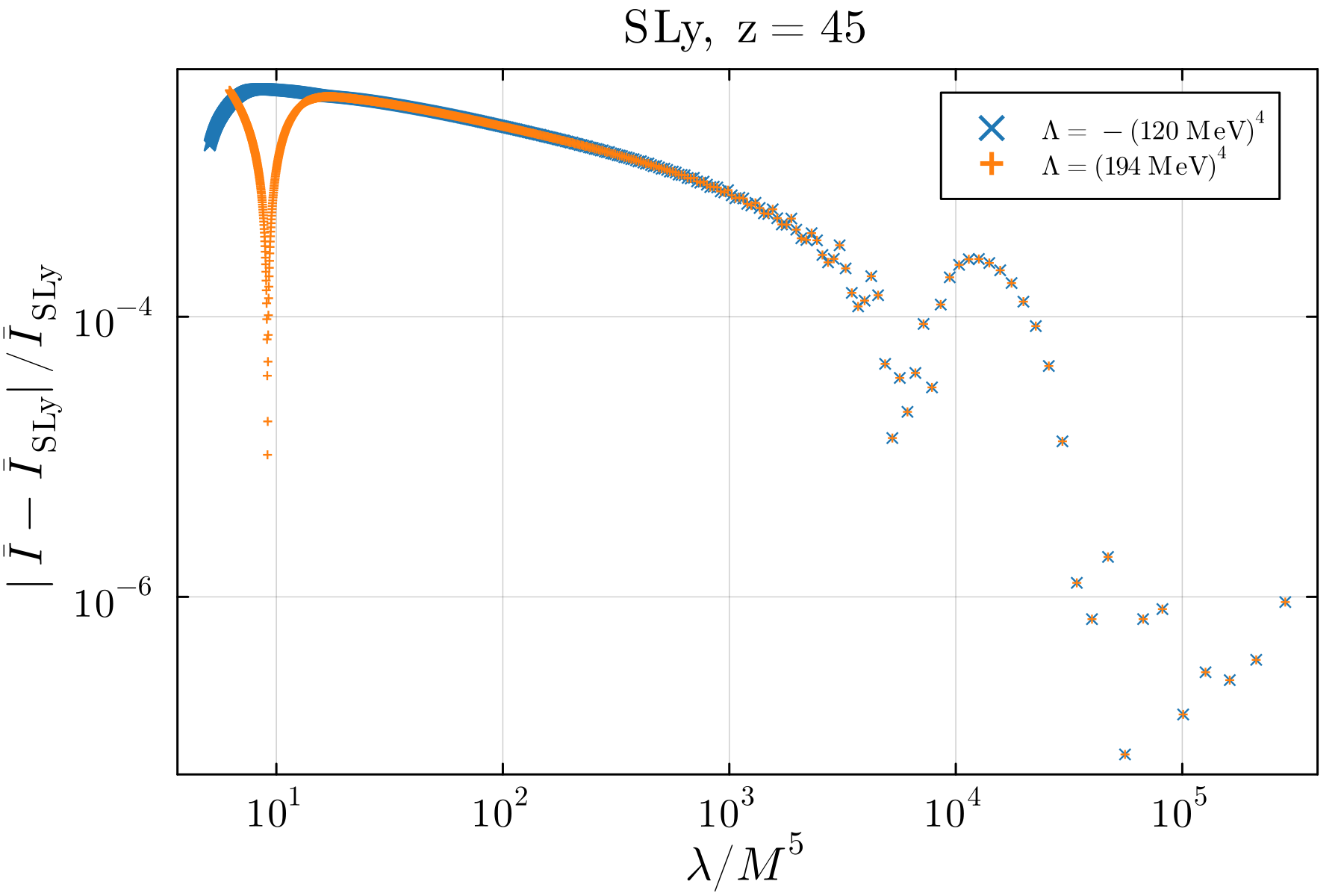}\hfill
\includegraphics[width=0.5\linewidth]{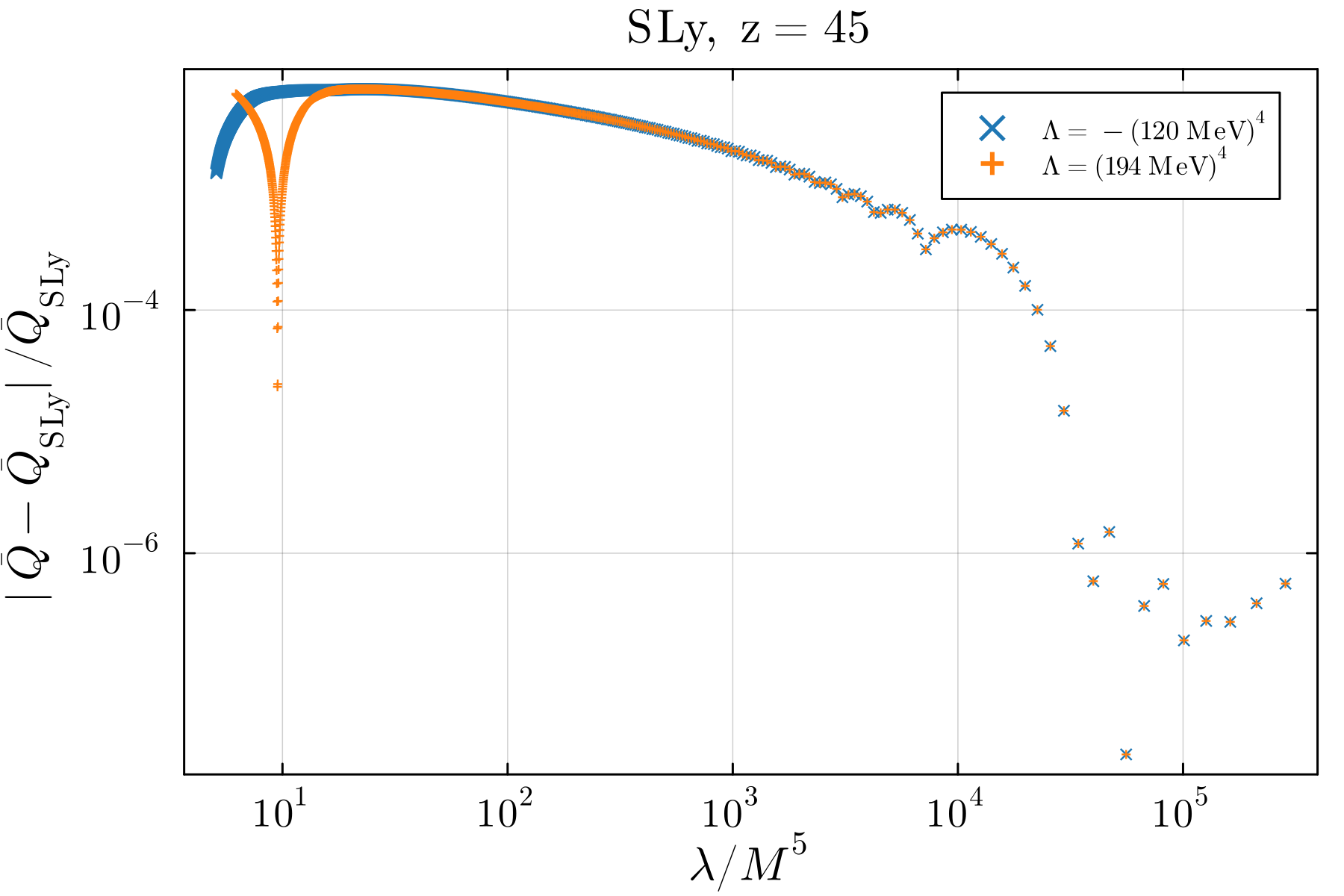}
	\caption{Fractional error in $Q$ and $I$ vs $\lambda$ for a specific high-density EOS modification, for slowly rotating neutron stars. The low-density description is taken to be the SLy EOS. We show the two cases of the most positive and most negative vacuum energy shifts allowed, i.e. $\Lambda= (194\,\text{MeV})^4$ and $\Lambda= -(120\,\text{MeV})^4$. The moment of inertia and the spin induced quadrupole moment have been appropriately normalized to $M^3$ and $M^3\chi^2$ respectively.}
	\label{fig:QIL45_SlowRot_Error}
\end{center}
\end{figure*}

We first consider the case of slowly rotating stars. We focus on the specific baseline EOS corresponding to label $z=45$ in Fig.~\ref{fig:fulldataset}, but the results generally hold for all possible configurations~\footnote{The results hold for EOSs belonging to both families we identified in the previous sections.}. In Fig.~\ref{fig:IQ45_SlowRot}, we show the $I$ vs $Q$ relation, where the quantities have been normalized to $M^3$ and $M^3\chi^2$ respectively, where $\chi=S/M^2$ is the dimensionless spin parameter. The left (right) panel corresponds to EOSs with the AP4 (SLy) as the low-density description, and we consider all possible values of vacuum energy jump admitted by the configuration. Similarly, we show the $I$ vs $\lambda$ and $Q$ vs $\lambda$ relations, where the tidal deformability has been normalized to $M^5$, in Figs.~\ref{fig:IL45_SlowRot} and~\ref{fig:QL45_SlowRot} respectively. In all cases, the deviations introduced by the new EOS prescription are barely perceptible. To better prove this point, we focus on the SLy case, and in Fig.~\ref{fig:QIL45_SlowRot_Error} we plot the fractional error $|I-I_\text{SLy}|/(I_\text{SLy})$ and $|Q-Q_\text{SLy}|/(Q_\text{SLy})$ versus the tidal deformability, where a subscript SLy denote a quantity for the standard SLy EOS. We show the two cases of the most positive and most negative vacuum energy shifts allowed, that is $\Lambda= (194\,\text{MeV})^4$ and $\Lambda= -(120\,\text{MeV})^4$, which introduce the largest corrections. Note that the curve dropping to zero is just a consequence of employing the absolute value of the difference $\Delta I$ and $\Delta Q$, which can change sign. For large values of the tidal deformability the deviations with respect to the standard SLy are around $0.01\%$ or less. For smaller values of normalized $\lambda$, the corrections increase and can reach up to $0.1\%$ in the case of very positive vacuum energy shift. These estimates are in agreement with current results on I-Love-Q relations~\cite{Carson:2019rjx}, and show that these universal relations still hold when introducing a vacuum energy transition in the core of a star.

\begin{figure*}[t!]
\begin{center}	\includegraphics[width=0.5\linewidth]{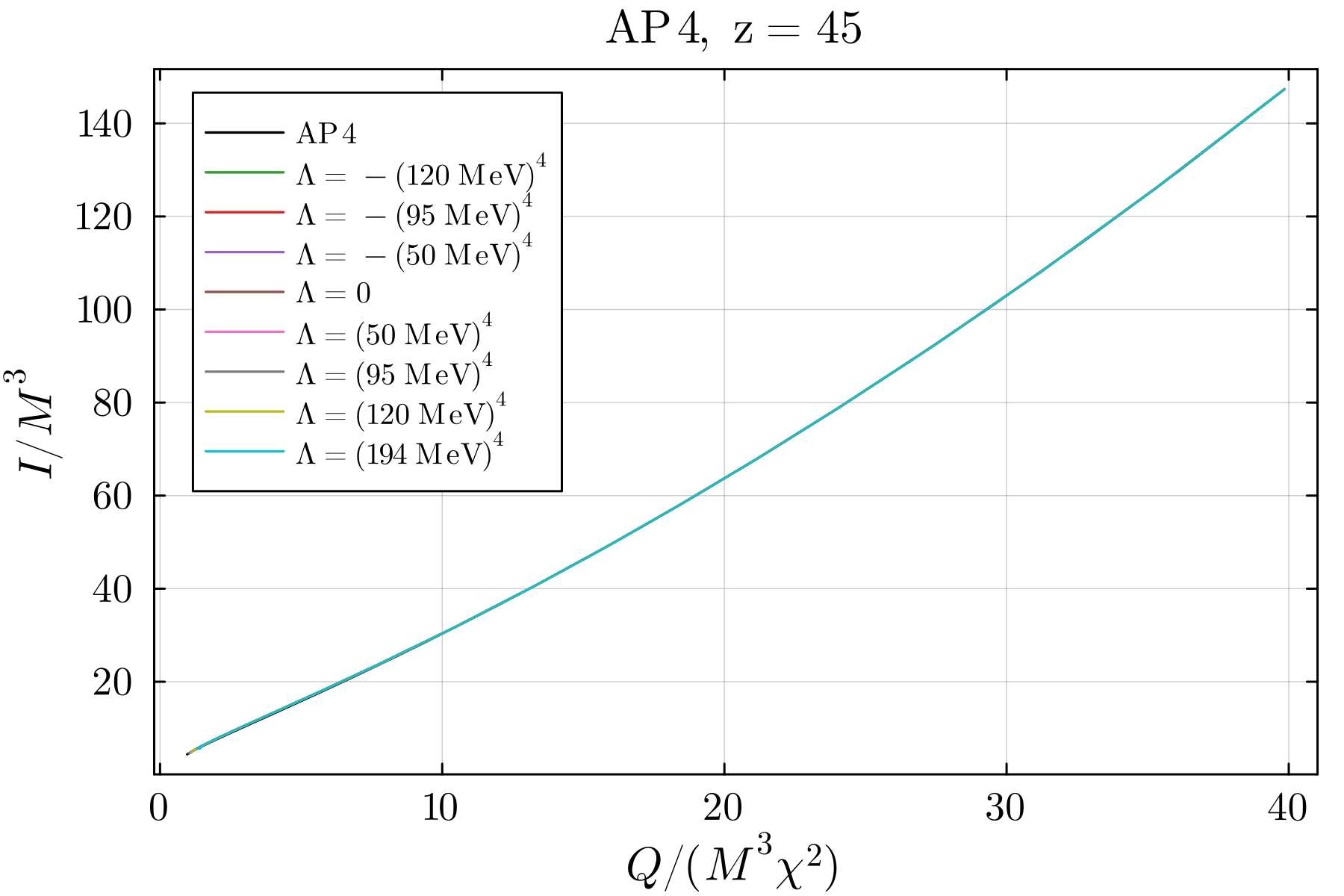}\hfill
\includegraphics[width=0.5\linewidth]{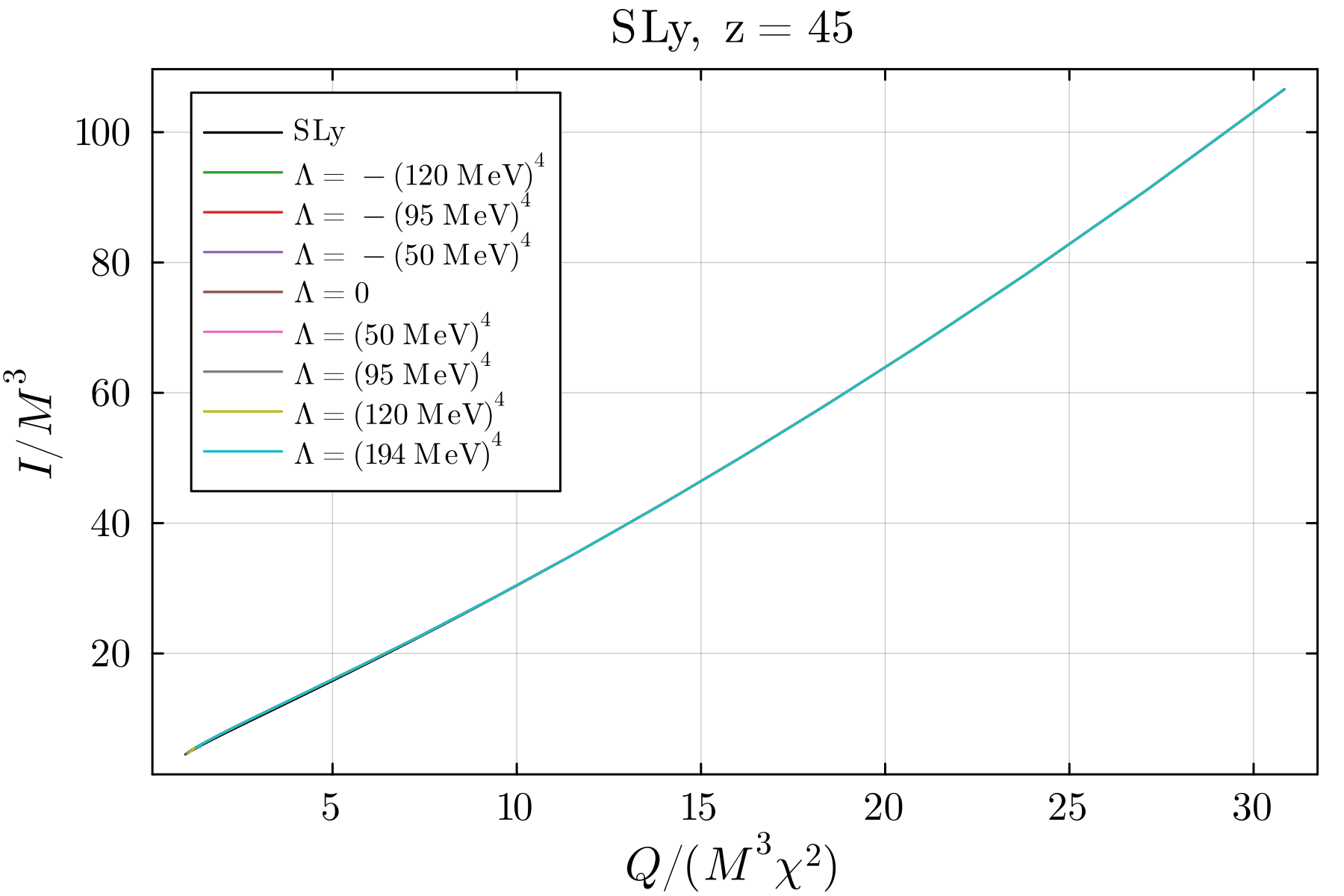}
	\caption{$Q$ vs $I$ for a specific high-density EOS modification, for maximally rotating neutron stars. The left (right) panel corresponds to the AP4 (SLy) as low-density description. We show the results for different values of vacuum energy shift, where differences are barely perceptible.}
	\label{fig:IQ45_Rot}
\end{center}
\end{figure*}

We now move on to consider fast rotating stars. Once again, we focus on the specific baseline EOS corresponding to label $z=45$ in Fig.~\ref{fig:fulldataset}, but the results generally hold. In Fig.~\ref{fig:IQ45_Rot} we plot the normalized moment of inertia $I/M^3$ as a function of the normalized quadrupole moment $Q/M^3$, for all values of $\Lambda$ allowed, for both AP4 (left) and SLy (right) low-density description. As in the case for slowly rotating stars, we observe that deviations between the different values of $\Lambda$ are barely perceptible (generically less than $\sim 1\%$) and only perceptible with respect to the vanilla EOSs without the QCD phase transition.

Based on these results, we can conclude that the I-Love-Q relations still hold when considering more exotic description in the core of the star, and are insensitive to the presence (or absence) of any new physics screening the cosmological constant in the particle physics sector.

\section{Conclusions}\label{sec:conc}

We have explored neutron stars as new probes for gravitational effects of the vacuum energy. Working in the context of General Relativity, we have modelled the interior of the star by employing an EOS composed of three regions, described by different approaches. At low density we used nucleonic tabulated EOSs (SLy and AP4); in the high-density regime, when the local density exceeds twice the nuclear saturation density, we modeled the QCD phase using an agnostic speed-of-sound parametrization; finally, when the pressure exceeds the QCD scale, we introduced a vacuum energy phase transition, allowing for both positive and negative values of its shift.

We have discovered that a vacuum energy phase transition in the core of the star impacts its mass-radius relations: a positive jump lowers the maximum mass, while a negative shift allows to reach for higher masses. This feature is different from the results previously obtained in~\cite{Csaki:2018fls}, where the maximum mass decreased for all possible values of vacuum energy phase transition. This is because we allow the total energy density to decrease in the core when the vacuum energy phase transition is triggered. In contrast, in~\cite{Csaki:2018fls}, the total energy density is always assumed to increase. We have also observed that the behaviour for rotating stars is qualitatively the same, the only difference being that same masses can allow for bigger radii, as expected when rotation kicks in.

Interestingly, we have discovered that the presence of the vacuum energy splits our EOSs into two families; on one side we found EOSs that permit a wide range of values for the cosmological constant, including $\Lambda=0$, on the other we obtained EOSs that are allowed only if a negative jump in vacuum energy is introduced. This is because implementing a large negative shift in vacuum energy in the core produces a sufficiently high maximum mass, thus ‘saving’ the configuration which otherwise would have been in disagreement with current observations~\cite{Romani:2022jhd}. We plan to extend our study to a wider range of values of the cosmological constant for future work. This will allow us to determine the threshold on $\Lambda$ between the two families. However, we expect that the threshold value of $\Lambda$ will slightly vary depending on the baseline EOS used to model the low-density region.

We have also investigated the effects of the new exotic EOSs on the tidal deformability. We have shown that EOSs belonging to the second family mentioned above exhibit a different tidal deformability-radius relation with respect to the first group of EOSs. This seems to confirm how these are different families of EOS, developing different properties. Focusing on observables from gravitational waves observation, we have demonstrated that the presence of both the QCD and the vacuum energy phase transition impacts on the combined tidal deformability for neutron stars binaries. Nonetheless, our results are still in agreement with current gravitational waves observation, that is $\tilde{\lambda}\leq 800$ for a chirp mass of $\mathcal{M}=1.188$ for all cases we considered.

Finally, we have shown that the I-Love-Q universal relations still holds when considering these new exotic EOSs. We have investigated both the case of slowly and fast rotating neutron stars, finding that the relations hold with a $\sim 1\%$ accuracy. We leave the study of other multipole moments universal relations~\cite{Pappas:2013naa,Stein:2013ofa,Yagi:2014bxa}, as well as universal relations for masses and radii~\cite{Konstantinou:2022vkr,Kruger:2023olj}, for future work.

We intend to constrain further our EOSs employing NICER measurements of masses and radii~\cite{Miller:2019cac,Riley:2019yda,Miller:2021qha,Riley:2021pdl} in a future work. Moreover, we plan to assess the detectability of the QCD and of the vacuum energy phase transition exploiting observations of binary neutron stars, in a follow-up work. In particular, we expect that gravitational wave observations from next-generation ground based detectors~\cite{Branchesi:2023mws}, as well as from electromagnetic facilities~\cite{Piro:2021oaa}, will put tighter constraints on Love numbers and on the stellar radii, allowing to identify the particle content of the neutron star dense matter~\cite{Pacilio:2021jmq}.

In this analysis, our focus has been on identifying signatures of a vacuum energy phase transition in the core of the star within the framework of General Relativity. Based on a canonical understanding of particle physics, such transitions are expected to occur whenever the pressure in the core reaches sufficiently large values.  In this paper, we have modelled the possibility that some new  physics alters the scale of that transition in some way, perhaps rendering it absent, screening the cosmological constant altogether,  or pushing it in some unexpected direction. As we have taken gravity to be consistently described by General Relativity, the assumption is that this new physics is non-gravitational. Of course, it could be that the screening of the cosmological constant is actually a gravitational phenomenon. How do we tell the difference? The key is to repeat our analysis for suitable modified theories of gravity and compare, focusing on I-Love-Q and mass-radius relations. 

To make direct contact with the cosmological constant problem, we should focus on modified gravity models that exhibit some degree of self-tuning. This allows us to directly compare the following two scenarios: General Relativity with the vacuum energy phase transition absent due to some non-gravitational screening mechanism;  and a self tuning theory, with a vacuum energy phase present present, but its effect screened by some gravitational dynamics.  Of course, the challenge is to find a suitable self tuning model. As a toy model of self-tuning, the Fab Four  \cite{Charmousis:2011bf,Charmousis:2011ea,Copeland:2012qf} could be a good place to start, although it is known that they are challenged by multi-messenger gravitational wave data \cite{Sakstein:2017xjx, Ezquiaga:2017ekz,Creminelli:2017sry,Baker:2017hug}, (see also, \cite{Langlois:2017dyl, Crisostomi:2017lbg, Babichev:2017lmw,Copeland:2018yuh,Kase:2018aps,Bordin:2020fww}). 

\section*{Acknowledgments}
We thank Ingo Tews, Miguel Bezares and Filippo Anzuini for useful discussions.
GV received support from the Czech Grant Agency (GA\^{C}R) under grant number 21-16583M.
PF is supported by a grant from Villum Fonden under Grant No.~29405, acknowledges support by a Research Leadership Award from the Leverhulme Trust and is funded by the Deutsche Forschungsgemeinschaft (DFG, German Research Foundation) under Germany’s Excellence Strategy EXC 2181/1 - 390900948 (the Heidelberg STRUCTURES Excellence Cluster). AM acknowledges financial support from MUR PRIN Grant No. 2022-Z9X4XS, funded by the European Union - Next Generation EU. AP and TS acknowledge partial support from the STFC Consolidated Grant nos. ST/V005596/1, ST/T000732/1, and ST/X000672/1.  For the purpose of open access, the authors have applied a CC BY public copyright licence to any Author Accepted Manuscript version arising. 

\appendix
\section{Convexity of the free energy}\label{App:convex}

We show here the conditions for the convexity of the free energy for the fluid, i.e. $\left( \frac{\partial^2F_\text{fl}}{\partial V^2} \right)_\text{T, N} > 0$, after the vacuum energy phase transition.

The Helmholtz free energy of the fluid is defined by $F_\text{fl} = U_\text{fl} - TS$, where $U_\text{fl}$ is the internal energy of the fluid, $T$ is the absolute temperature and $S$ is the entropy. Let T and N, the number of particles, be fixed, we then have
\be\label{eq:freeEn}
    \text{d}F_\text{fl} = \text{d}U_\text{fl} - \text{d}T S - T \text{d}S = T \text{d}S - p_\text{fl} \text{d}V - T \text{d}S.
\ee
Using Eq.~\eqref{eq:freeEn}, we find
\be
    \left. \frac{\partial F_\text{fl}}{\partial V} \right\rvert_\text{T, N} = -p_\text{fl} \, \rightarrow \, \left. \frac{\partial^2 F_\text{fl}}{\partial V^2} \right\rvert_\text{T, N} = - \left. \frac{\partial p_\text{fl}}{\partial V} \right\rvert_\text{T, N},
\ee
and the convexity of the free energy requires $\partial p_\text{fl}/\partial V < 0$, or alternatively one finds $\partial p_\text{fl}/\partial \rho > 0$. Hence, the fluid pressure must increase (decrease) with the mass density. Note that, at the phase transition junction, the pressure of the fluid undergoes a jump proportional to $\Lambda$ since one has
\be
    \Delta p = p_\text{in} - p_\text{out} = p_c + \Lambda - p_c = \Lambda.
\ee
Now, we are faced with two possible scenarios. If the transition generates a positive cosmological constant $\Lambda>0$, the fluid pressure in the interior must also increase. This in turns requires a positive jump in the total mass density and, consequently, total energy density from Eq.~\eqref{eq:condEn}. If the transition generates a negative cosmological constant $\Lambda < 0$, the fluid pressure will then decrease, leading to a drop in total mass density and also total energy density.

\section{Exterior solution for slowly rotating stars at second order}\label{App:Slowly}

The exterior solutions to the system of Eqs.~\eqref{eq:slowly2.1}-\eqref{eq:slowlydT} obtained by imposing asymptotic flatness at spatial infinity are given by~\cite{Hartle:1967he}
\begin{widetext}
    \begin{align}
        h^\text{ext}_2 & = \frac{1}{M r^3}\left( 1+\frac{M}{r} \right)S^2 + A Q^2_2(r/M-1) \nonumber \\
        &= \frac{1}{M r^3}\left( 1+\frac{M}{r} \right)S^2 -\frac{3 A r^2}{M(r-2 M)}\left[ 1-3\frac{M}{r}+\frac{4}{3}\frac{M^2}{r^2}+\frac{2}{3}\frac{M^3}{r^3}+\frac{r}{2 M}f(r)^2 \text{ln}f(r)\right]\ , \\
        K^\text{ext}_2 & = -\frac{1}{M r^3}\left( 1+\frac{M}{r} \right)S^2 + \frac{2A M}{\sqrt{r(r-2M)}}Q^1_2(r/M-1)-AQ^2_2(r/M-1) \nonumber \\
        & = -\frac{1}{M r^3}\left( 1+\frac{M}{r} \right)S^2 + \frac{3Ar}{M}\left[ 1 + \frac{M}{r} - \frac{2}{3} \frac{M^2}{r^2} +\frac{r}{2M}\left( 1-\frac{2M^2}{r^2} \right)\text{ln}f(r) \right]\ , \\
        m^\text{ext}_2 & = - \frac{1}{M r^2}\left( 1-7\frac{M}{r}+10 \frac{M^2}{r^2} \right) S^2 +\frac{3 A r^2}{M} \left[ 1-3\frac{M}{r}+\frac{4}{3}\frac{M^2}{r^2}+\frac{2}{3}\frac{M^3}{r^3}+\frac{r}{2M}f(r)^2 \text{ln}f(r) \right]\ ,
    \end{align}
\end{widetext}
where$f(r)= (1-2M/r)$, $Q^1_2$ and $Q^2_2$ are the associated Legendre function of the second kind and A is an integration constant to be determined by the matching of the interior solution to the exterior one at the surface.

\section{Source terms for rotating stars}\label{Appendix:Rotating}

The source terms appearing in Eqs. \eqref{Eq:FEq1}, \eqref{Eq:FEq2}, \eqref{Eq:FEq3} and \eqref{Eq:FEq4} are given by
\begin{widetext}
\begin{equation}
    \begin{aligned}
    S_\rho(r,\theta) =  e^{\gamma/2} \bigg[& 8\pi e^{2\alpha} \left(\epsilon + p\right) \frac{1+v^2}{1-v^2} + r^2 \sin^2 \theta e^{-2\rho} \left( \omega_{,r}^2 + \frac{1}{r^2}\omega_{,\theta}^2 \right) \\&
    + \frac{1}{r} \gamma_{,r} + \frac{1}{r} \cot \theta \gamma_{,\theta} + \frac{\rho}{2} \bigg\{ 16\pi e^{2\alpha} p -\gamma_{,r} \left(\frac{1}{2}\gamma_{,r} + \frac{1}{r}\right) 
    -\frac{1}{r^2} \gamma_{,\theta} \left(\frac{1}{2} \gamma_{,\theta} + \cot \theta\right)
    \bigg\}
    \bigg],
    \end{aligned}
\end{equation}
\begin{equation}
    \begin{aligned}
    S_\omega(r,\theta) =  e^{(\gamma-2\rho)/2} \bigg[& -16\pi e^{2\alpha} \frac{(\Omega-\omega)(\epsilon+p)}{1-v^2} + \omega \bigg\{ -8\pi e^{2\alpha} \frac{(1+v^2)\epsilon+2v^2p}{1-v^2} - \frac{1}{r} \left(2 \rho_{,r} + \frac{1}{2} \gamma_{,r}\right) \\&
    \frac{\cot \theta}{r^2} \left(2\rho_{,\theta} +\frac{1}{2} \gamma_{,\theta}\right) + \rho_{,r}^2 -\frac{1}{4} \gamma_{,r}^2 + \frac{1}{r^2} \left(\rho_{,\theta}^2 -\frac{1}{4} \gamma_{,\theta}^2 - r^2 \sin^2 \theta e^{-2\rho} \left[\omega_{,r}^2 + \frac{1}{r^2} \omega_{\theta} ^2\right]\right)
    \bigg\}
    \bigg],
    \end{aligned}
\end{equation}
\begin{equation}
    \begin{aligned}
    S_\gamma(r,\theta) =  e^{\gamma/2} \bigg[& 16\pi e^{2\alpha} p + \frac{\gamma}{2} \left(16\pi e^{2\alpha} p - \frac{1}{2} \left(\gamma_{,r}^2 + \frac{1}{r^2} \gamma_{,\theta}^2\right)\right)
    \bigg],
    \end{aligned}
\end{equation}
\begin{equation}
    \begin{aligned}
    S_\alpha(r,\theta) =&  -\frac{1}{2}(\rho_{,\mu}+\gamma_{,\mu})+\bigg\{\frac{1}{2}[r^2(\gamma_{,rr}+\gamma^2_{,r})-(1-\mu^2)(\gamma_{,\mu\mu}+\gamma^2_{,\mu})][-\mu+(1-\mu^2)\gamma_{,\mu}] \\&
+r\gamma_{,r}[\frac{1}{2}\mu+\mu r \gamma_{,r}+\frac{1}{2}(1-\mu^2)\gamma_{,\mu}]+\frac{3}{2}\gamma_{,\mu}[-\mu^2+\mu(1-\mu^2)\gamma_{,\mu}] \\&
-r(1+r \gamma_{,r})(1-\mu^2)(\gamma_{,r\mu}+\gamma_{,r}\gamma_{,\mu})-\frac{1}{4}\mu r^2 (\rho_{,r}+\gamma_{,r})^2-\frac{1}{2}r (1+r\gamma_{,r})(1-\mu^2)(\rho_{,r}+\gamma_{,r})(\rho_{,\mu}+\gamma_{,\mu}) \\&
+\frac{1}{4}\mu (1-\mu^2)(\rho_{,\mu}+\gamma_{,\mu})^2+\frac{1}{4}r^2 \mu (1-\mu^2)\gamma_{,\mu}[r^2(\rho_{,r}+\gamma_{,r})^2-(1-\mu^2)(\rho_{,\mu}+\gamma_{,\mu})^2] \\&
+(1-\mu^2) e^{-2\rho}(\frac{1}{4}r^4\mu \omega^2_{,r}+\frac{1}{2}r^3(1-\mu^2) \omega_{,r}\omega_{,\mu}-\frac{1}{4}r^2 \mu(1-\mu^2) \omega^2_{,\mu}+\frac{1}{2}r^4(1-\mu^2)\gamma_{,r}\omega_{,r}\omega_{,\mu} \\&
-\frac{1}{4} r^2(1-\mu^2)\gamma_{,\mu}[r^2 \omega^2_{,r}-(1-\mu^2)\omega^2_{,\mu}])\}/\{(1-\mu^2)(1+r\gamma_{,r})^2+[\mu-(1-\mu^2)\gamma_{,\mu}]^2\bigg\}\ ,
    \end{aligned}
\end{equation}
where we have used $\mu=\cos \theta$ as an independent variable to avoid the apparent singularity of $\cot \theta$ at the pole.

The expansions of the Green's functions in angular harmonics are given by
\begin{equation}
    \frac{1}{|\mathbf{r}-\mathbf{r'}|} = \sum_{n=0}^\infty \frac{r_<^n}{r_>^{n+1}} \left[ P_n(\cos \theta) P_n(\cos \theta') + 2\sum_{m=1}^n \frac{(n-m)!}{(n+m)!} P^m_n(\cos \theta) P^m_n(\cos \theta') \cos \left\{ m (\phi-\phi') \right\} \right],
\end{equation}
\begin{equation}
    \log |\mathbf{r}-\mathbf{r'}| = -\sum_{n=1}^\infty \frac{1}{n} \frac{r_<^n}{r_>^{n}} \left[ \cos n \theta \cos n \theta' + \sin n \theta \sin n \theta' \right] + \log(r_>),
\end{equation}
where $r_< = \mathrm{min}(r,r')$, and $r_> = \mathrm{max}(r,r')$, and $P_n$ and $P^m_n$ denote the Legendre and associated Legendre polynomials, respectively.
\end{widetext}

\section{Relative shift in combined dimensionless tidal deformability}\label{Appendix:RelativeShift}

\begin{figure*}[h!]
\begin{center}
 \subfloat{%
	\includegraphics[width=0.5\linewidth]{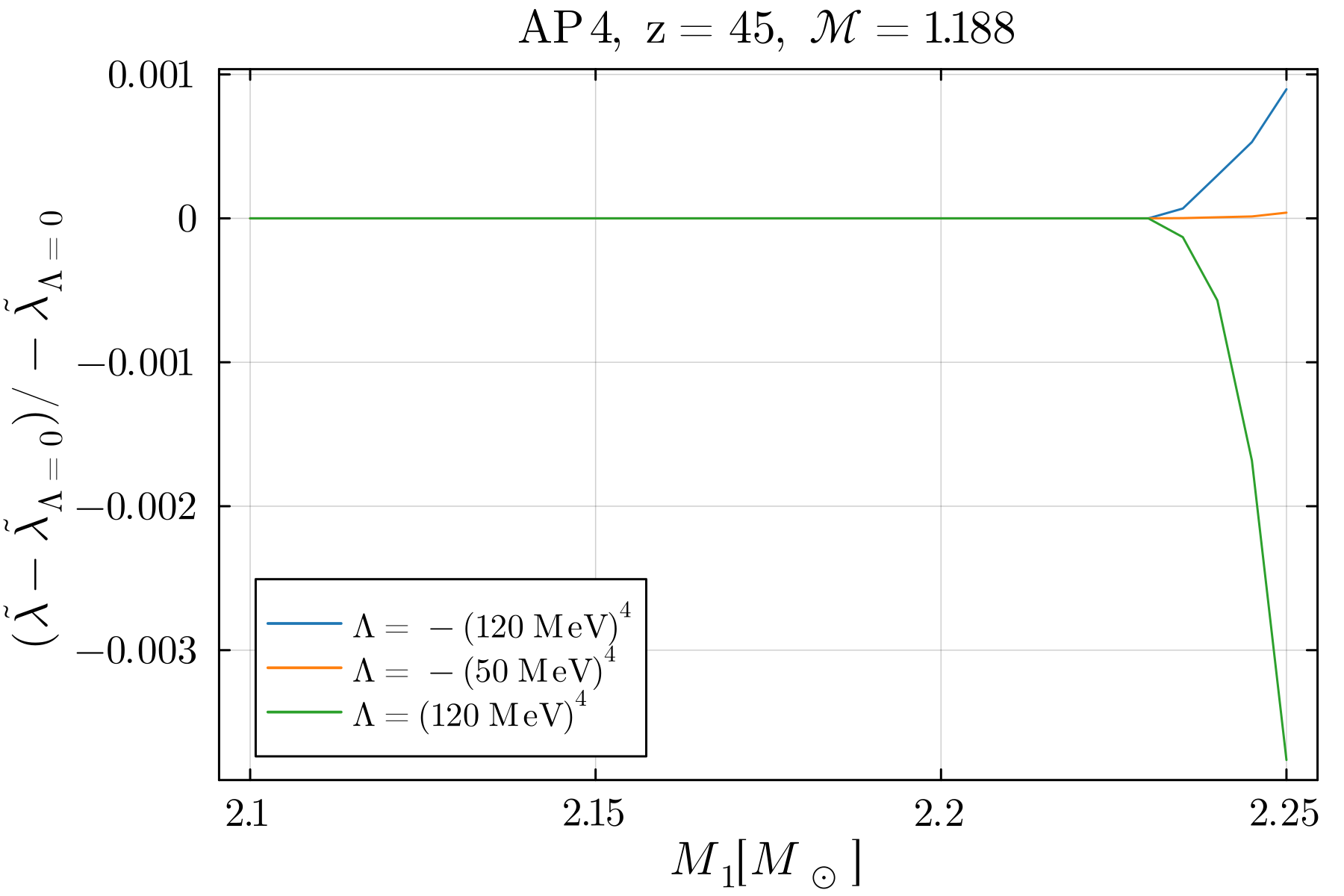}%
	}
	\subfloat{%
	\includegraphics[width=0.5\linewidth]{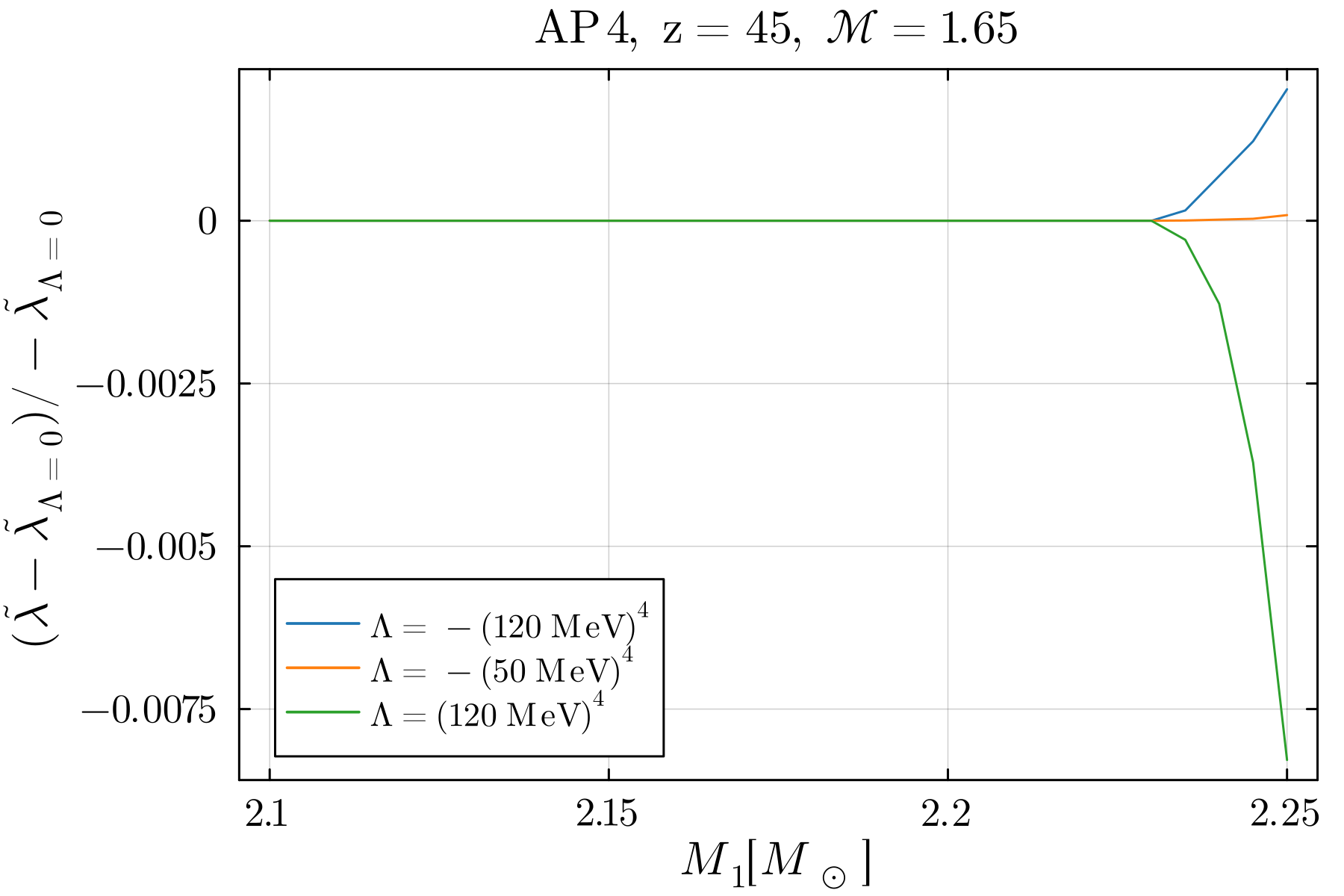}%
	}
	\caption{Relative shift in combined dimensionless tidal deformability $\tilde{\lambda}$ for EOSs built from the standard AP4 corresponding to a specific high-density EOS modification. We compare EOSs with different values of $\Lambda$ with the baseline EOS (with $\Lambda=0$).}
	\label{fig:shifts}
\end{center}
\end{figure*}

Here we show the results for the relative shift in combined dimensionless tidal deformability $\tilde{\lambda}$ for the EOSs built from the standard AP4 considered in section~\ref{sec:tidalDef}. We show the results when comparing the baseline EOS corresponding to label $z=45$ in Fig.~\ref{fig:fulldataset} with different values of $\Lambda$ with the baseline EOS with no vacuum energy contribution (with $\Lambda=0$).

%
\bibliography{Ref}

\end{document}